\documentclass[twocolumn]{aastex701}

\usepackage{amsmath}
\usepackage{xcolor}

\usepackage{soul}
\soulregister\citep7
\soulregister\citet7
\soulregister\citealt7
\soulregister\ref7
\soulregister\cite7
\shortauthors{Zhao, Zhang, \& Zhang}

\begin{document}

\title{Gamma-Ray Emission from LS~5039: A 30-500~GeV Excess and Phase-Resolved Broadband Spectral Modeling}

%\correspondingauthor{Pengfei Zhang}
%\email{zhangpengfei@ynu.edu.cn}

\author[0009-0008-3339-2035]{Xiaozhen Zhao}
\affiliation{Department of Astronomy, School of Physics and Astronomy, Key Laboratory of Astroparticle Physics of Yunnan Province, Yunnan University, Kunming 650091, People's Republic of China; zhangpengfei@ynu.edu.cn}
\email{zhangpengfei@ynu.edu.cn}

\author{Jiandong Zhang}
\affiliation{Department of Astronomy, School of Physics and Astronomy, Key Laboratory of Astroparticle Physics of Yunnan Province, Yunnan University, Kunming 650091, People's Republic of China; zhangpengfei@ynu.edu.cn}
\email{zhangpengfei@ynu.edu.cn}

\author{Pengfei Zhang}
\affiliation{Department of Astronomy, School of Physics and Astronomy, Key Laboratory of Astroparticle Physics of Yunnan Province, Yunnan University, Kunming 650091, People's Republic of China; zhangpengfei@ynu.edu.cn}
\email{zhangpengfei@ynu.edu.cn}

\begin{abstract}

LS~5039 is a well-studied $\gamma$-ray binary exhibiting strong orbital
modulation from GeV to TeV energies. We present a phase-resolved analysis
of 17.3~yr of \textit{Fermi}-LAT observations and
identify an additional $\gamma$-ray excess in the $\sim30$--$500$~GeV
at a significance of $\sim7.9\sigma$.
The phase-resolved spectra show pronounced orbital dependence above
$\sim10$~GeV, with the inferior-conjunction more complex than the superior-conjunction.
To investigate the broadband spectral energy distribution, we combine the
\textit{Fermi}-LAT spectra with published H.E.S.S.\ and HAWC measurements
and compare purely leptonic and lepto-hadronic models using the Bayesian
information criterion.
In both intervals, relativistic-Maxwellian and magnetospheric
components dominate below $\sim10$~GeV, while shock-accelerated
electrons account for much of the GeV--TeV continuum.
The purely leptonic fits leave residuals at $30$--$500$~GeV and
underpredict the HAWC measurements above $\sim100$~TeV. In the local
one-zone intrabinary shock apex model, synchrotron cooling and Klein-Nishina
suppression make the ultra-high-energy emission difficult to reproduce, although
more complex leptonic scenarios remain possible.
Adding a phenomenological proton component improves the broadband
description, with stronger relative support during superior-conjunction than inferior-conjunction.
However, the proton parameters remain weakly constrained and the
fast-escape hadronic interpretation is energetically challenging, so
a hadronic origin is not uniquely established.
Future high-sensitivity, phase-resolved observations at multi-TeV and
ultra-high energies will be required to test this interpretation.

\end{abstract}


\section{Introduction}
\label{sec:in}

Gamma-ray binaries are compact binary systems composed of a massive star and a compact object,
in which most of the nonthermal radiative output is emitted at $\gamma$-ray energies,
above $\sim1$~MeV. Their high-energy radiation is generally attributed to particles accelerated
in relativistic outflows or wind-interaction regions associated with the compact object and
its massive companion. The observed emission can involve synchrotron radiation,
anisotropic inverse-Compton (IC) scattering, photon--photon pair production
($\gamma\gamma\rightarrow e^{+}e^{-}$) and the resulting absorption and pair cascading,
as well as potentially hadronic processes \citep{dubus13}. Orbital motion periodically changes
the viewing geometry, stellar-photon and wind densities, and $\gamma\gamma$ opacity.
These systems therefore provide unique laboratories for studying extreme particle acceleration
and radiative processes under periodically varying physical conditions.

The origin of the high-energy emission from $\gamma$-ray binaries
remains debated. In the microquasar scenario, accretion powers
relativistic jets that can produce leptonic and hadronic emission
\citep{mr99,bp04,db06,zzf09}. In the pulsar-wind scenario, a
rotation-powered neutron-star wind collides with the stellar wind,
forming an intrabinary shock (IBS) where particles can be accelerated
\citep{ta97,gs06,st08}. Shock-accelerated electrons can produce
synchrotron and IC emission, while relativistic protons may generate
$\gamma$-rays through $pp$ interactions. Magnetospheric outer-gap
emission \citep{zc97,crz00,tlt+14,cyz24} and other pulsar-wind
processes, including free-wind emission, pair cascades, and striped-wind
dissipation \citep{cmd+10,pd11}, may also contribute. Their different
emission sites and phase dependences make phase-resolved observations
essential for distinguishing these scenarios.

To date, only a handful of $\gamma$-ray binaries have been firmly identified in the
Galaxy \citep{aab+06,aaa+06,hsf+09,ccc+2019}. LS~5039 is among the most prominent and
best-characterized members of this class. It consists of an O6.5\,V((f)) star with
a mass of $M_\star\approx23\,M_\odot$ and a compact companion in an eccentric orbit with
a period of $\sim3.9$~d and an eccentricity of $e\approx0.35$ \citep{crr+05,msa+16}.
Although the nature of the compact object remains uncertain, the reported detection of
a $\sim9$~s periodic signal in hard X-rays provides support for a neutron-star
interpretation \citep{yme+20}. In this work, we adopt the neutron-star and pulsar-wind
scenario as a working assumption and investigate its implications for the high-energy emission.
With its well-determined orbital ephemeris, extensive multiwavelength coverage, and recently
reported ultra-high-energy (UHE) measurements, LS~5039 provides an exceptional laboratory for
testing particle-acceleration and radiation models in $\gamma$-ray binaries. In particular,
the available GeV--UHE observations span approximately five decades in energy,
enabling phase-resolved tests of the underlying particle populations and emission mechanisms.

LS~5039 exhibits nonthermal emission from radio to TeV energies, with clear orbital modulation
observed across multiple wavebands \citep{prm00,aaa+05,crr+05,aaa+09,tku+09,cz14}.
H.E.S.S. observations above $0.1$~TeV revealed a hard TeV spectrum with
strong phase-dependent behavior, whereas \textit{Fermi}-LAT observations revealed
a bright GeV component with a spectral cutoff at a few GeV and a distinct modulation
pattern \citep{aab+06,aaa+09}. This GeV--TeV dichotomy suggests that the observed emission
cannot be described by a single simple radiative component and may require
multiple particle populations, emission zones, or phase-dependent acceleration and
cooling processes. The transition region between the GeV and TeV components is therefore
particularly important for understanding the radiation mechanisms in LS~5039.

Recent HAWC observations have further extended the spectrum of LS~5039 into
the UHE regime, with $\gamma$-rays detected up to $\sim200$~TeV \citep{aaa+25}.
These measurements provide important additional constraints on
broadband emission models. In the IBS environment, radiative cooling and
the Klein--Nishina suppression of IC scattering can strongly limit the highest-energy IC
emission from relativistic electrons. The UHE observations therefore motivate testing whether
the GeV--TeV transition and the UHE spectral extension can be described within
a purely leptonic framework, or whether including a hadronic contribution provides a better
phenomenological description of the UHE emission.

In this work, we analyze 17.3~yr of \textit{Fermi}-LAT observations of LS~5039, focusing on
the high-energy end of the LAT spectrum, the high-energy excess around $30$--$500$~GeV,
and their orbital-phase dependence. We combine the phase-resolved LAT spectra with
published H.E.S.S. and HAWC measurements to model the broadband
spectral energy distribution (SED) within an IBS framework
and compare purely leptonic and lepto-hadronic scenarios. Section~\ref{sec:data} describes
the \textit{Fermi}-LAT analysis, Sections~\ref{sec:rapa} and \ref{sec:specfit} present
the physical model and fitting procedure, respectively, Section~\ref{sec:result} gives
the results and discussion, and Section~\ref{sec:sum} summarizes our conclusions.

%In this work, we perform an updated analysis of 17~yr of \textit{Fermi}-LAT observations of LS~5039, with particular attention to the high-energy end of the LAT spectrum, the additional $30$--$500$~GeV emission, and their orbital-phase dependence. We then combine the phase-resolved LAT spectra with published H.E.S.S. and HAWC measurements to construct and model the broadband SED. Our aim is to investigate the connection among the GeV cutoff component, the GeV--TeV transition, and the UHE $\gamma$-ray emission within a physically motivated IBS framework. To this end, we construct a broadband model that includes magnetospheric emission, leptonic IC emission from two electron populations, and a possible hadronic component produced through inelastic $pp$ interactions.

%The remainder of this paper is organized as follows. Section~\ref{sec:data} describes the data analysis of \textit{Fermi}-LAT and the construction of phase-resolved spectra. Section~\ref{sec:rapa} presents the physical framework, including the IBS geometry, radiative processes, particle cooling, and energy constraints. Section~\ref{sec:specfit} introduces the broadband SED model and fitting procedure. Section~\ref{sec:result} presents the phase-resolved fitting results, including the additional $30$--$500$~GeV emission and the comparison between the purely leptonic and lepto-hadronic scenarios. Finally, Section~\ref{sec:sum} summarizes our main conclusions.

\section{Fermi Observation and Data Reduction}
\label{sec:data}

\subsection{Data Reduction}
We analyzed $\gamma$-ray events collected by the \textit{Fermi}-LAT from 2008 August 4 to
2025 November 25 (MJD 54682.687--61004.000). The events were selected within
a $15^\circ \times 15^\circ$ region of interest (ROI) centered on 4FGL~J1826.2$-$1450 (LS~5039)
$(\mathrm{R.A.}=276^\circ.5637,\mathrm{decl.}=-14^\circ.8496)$. Our data analysis
based on the standard Pass~8 data-selection criteria with SOURCE-class Front+Back events (\texttt{evclass=128},
\texttt{evtype=3}) in the energy range from 0.1 to 500.0~GeV. To reduce the $\gamma$-ray contamination
from the Earth Limb, we applied a maximum zenith-angle cut of $90^\circ$. And the good time intervals were
selected using the standard filter expression \texttt{DATA\_QUAL>0 \&\& LAT\_CONFIG==1}. The data reduction
and likelihood analysis were carried out with \texttt{Fermitools} version of 2.2.0 with
the instrument response functions of P8R3\_SOURCE\_V3. The Galactic and extragalactic diffuse $\gamma$-ray
emissions were modeled using \texttt{gll\_iem\_v07.fits}
and the isotropic diffuse template \texttt{iso\_P8R3\_SOURCE\_V3\_v1.txt}.

\begin{deluxetable*}{ccccccc}
\tablecaption{Best-fit spectral parameters for the phase-averaged and phase-resolved spectra}
\tablewidth{0pt}
\tablehead{
    \colhead{Model} &
    \colhead{Parameter 1} &
    \colhead{Parameter 2} &
    \colhead{Characteristic energy (GeV)} &
    \colhead{TS} &
    \colhead{$-\ln\mathcal{L}_{\max}$} &
    \colhead{AIC}
}
\startdata
LP
& $\alpha=2.698(14)$
& $\beta=0.041(5)$
& $E_{\rm b}=1.991$
& 32956.0
& -44716260.45
& -89432078.90 \\
PLEC4
& $\Gamma=2.612(7)$
& $b=0.666$
& $E_{\rm c}=10.9(22)$
& 33031.2
& -44716259.44
& -89432076.88 \\
BPL
& $\Gamma_1=2.540(7)$
& $\Gamma_2=3.059(48)$
& $E_{\rm b}=2.321(73)$
& 33125.2
& -44716289.64
& -89432135.28 \\
\tableline
BPL$_{\textbf{SUPC}}$
& $\Gamma_1=2.406(12)$
& $\Gamma_2=2.908(27)$
& $E_{\rm b}=0.884(34)$
& 23611.2
& -19011191.34
& -- \\
BPL$_{\textbf{INFC}}$
& $\Gamma_1=2.644(18)$
& $\Gamma_2=2.590(31)$
& $E_{\rm b}=0.992(56)$
& 10111.2
& -13747005.83
& -- \\
\enddata
\tablecomments{Best-fit spectral parameters obtained from the
0.1--500.0~GeV spectral fits. The phase-averaged spectrum was
fitted with the LP, PLEC4, and BPL models, whereas the SUPC
and INFC spectra were fitted with the BPL model.
The characteristic energy is model dependent:
$E_{\rm b}$ is the pivot energy for LP and the break energy
for BPL, whereas $E_{\rm c}$ is the equivalent cutoff energy
for PLEC4. Numbers in parentheses denote the $1\sigma$
statistical uncertainties in the last quoted digits,
with uncertainties rounded to two significant figures.
Parameters without parenthetical uncertainties were fixed
during the fit. TS values should be compared only among
fits using the same data selection.}
\label{tab:paras}
\label{tab:orbitalpara}
\end{deluxetable*}

\begin{deluxetable*}{lccc}
\tablecaption{Best-fit parameters of the BPL+LP model for the phase-averaged, SUPC, and INFC spectra}
\tablewidth{0pt}
\tabletypesize{\scriptsize}
\tablehead{
\colhead{BPL+LP parameter}
& \colhead{All phase}
& \colhead{SUPC}
& \colhead{INFC}
}
\startdata
Parameter 1
& $\Gamma_1=2.541(7)$
& $\Gamma_1=2.503(9)$
& $\Gamma_1=2.587(13)$ \\
Parameter 2
& $\Gamma_2=3.86(17)$
& $\Gamma_2=3.58(10)$
& $\Gamma_2=3.29(16)$ \\
Characteristic energy 1 (GeV)
& $E_{\rm b}=3.52(11)$
& $E_{\rm b}=2.59(10)$
& $E_{\rm b}=3.54(20)$ \\
Parameter 3
& $\alpha=2.12(40)$
& $\alpha=2.23(31)$
& $\alpha=2.20(41)$ \\
Parameter 4 & $\beta=0.51(7)$ & $\beta=0.61(17)$ & $\beta=1.24(44)$ \\
Characteristic energy 2 (GeV) & $E_{\rm b,LP}=100$ & $E_{\rm b,LP}=100$ & $E_{\rm b,LP}=100$ \\
TS$_{\rm BPL}$ & 33023.4 & 24398.4 & 9137.6 \\
TS$_{\rm LP}$  & 116.9 & 39.6 & 50.5 \\
$-\ln\mathcal{L}_{\max}$
& $-44716325.81$ & $-19011195.91$ & $-13747011.55$
\enddata
\tablecomments{Parameters 1 and 2 and characteristic energy 1
describe the BPL component, while Parameters 3 and 4 and
characteristic energy 2 describe the additional LP component.
The LP reference energy was fixed at 100~GeV.
Numbers in parentheses denote the $1\sigma$ statistical
uncertainties in the last quoted digits, with uncertainties
rounded to two significant figures.
TS$_{\rm BPL}$ and TS$_{\rm LP}$ denote the component-wise
source-detection TS values reported for the BPL and
additional LP components, respectively; they should not
be confused with the likelihood-ratio $\Delta{\rm TS}$
obtained by comparing the complete BPL and BPL+LP models.}
\label{tab:bpl_lp_parameters}
\end{deluxetable*}

\subsection{Spectral Analysis}

Based on a model file generated
by make4FGLxml.py\footnote{https://fermi.gsfc.nasa.gov/ssc/data/analysis/user/}
that describes all $\gamma$-ray sources in 4FGL-DR4 within the ROI, we performed
a binned maximum  likelihood spectral analysis. In the file, the target source
LS~5039 has a log-parabola spectral shape (LP). The spectral shapes of other
point sources were taken from the 4FGL-DR4, and the spectral parameters of sources
located more than $10^\circ$ from the ROI center were fixed to their catalog values,
while others were set free. We tested other two alternative spectral models, a power
law with a super-exponential cutoff (\texttt{PLEC4}) and a broken power
law (\texttt{BPL}), to describe the $\gamma$-ray emissions from LS~5039.
Table~\ref{tab:paras} summarizes the best-fit parameters and
source-detection TS values for the three spectral models. Because LP,
PLEC4, and BPL have different numbers of free parameters and are not
all mutually nested, we compare them using the Akaike information
criterion (AIC; \citealt{Akaike74}),
${\rm AIC}=2k-2\ln\mathcal{L}_{\max}$. The BPL model has the lowest
AIC and is therefore adopted for the subsequent phase-resolved
analysis.

\begin{figure}
\centering
\includegraphics[angle=0,scale=0.52]{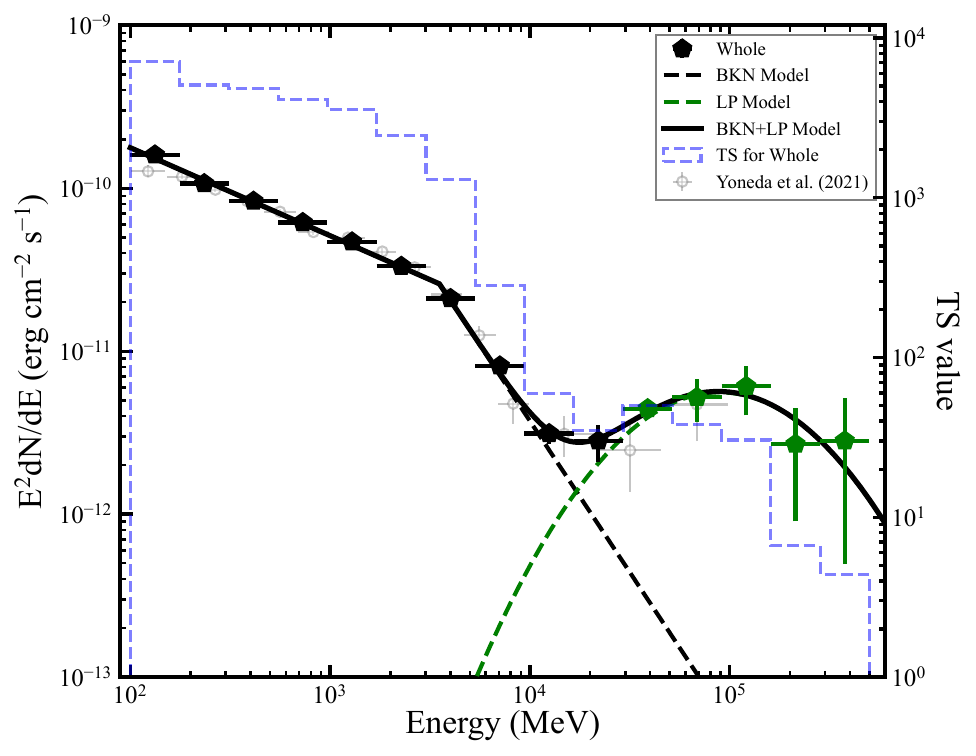}
\caption{SED of LS~5039 in the $0.1$--$500.0$~GeV range. Filled
pentagons show the measurements obtained in this work, with the green
points highlighting the high-energy excess at $\sim30$--$500$~GeV.
The black dashed and green dashed curves represent the BPL and
additional LP components, respectively, while the black solid curve
shows their combined BPL+LP model. The blue dashed step curve shows
the test statistic (TS) in each energy bin, with values given by the
right-hand y-axis. All SED data points satisfy $\mathrm{TS}>4$.
For comparison, the SED reported by \citet{yke+21} is shown as open
gray circles.}
\label{fig:allsed}
\end{figure}
%%%%%%%%%%%%

To characterize the $\gamma$-ray SED of LS~5039, we divided the
0.1--500.0~GeV energy range into 15 equally logarithmically spaced energy bins and performed
a binned maximum-likelihood analysis independently in each bin. In each fit, LS~5039 was
described by the best-fit BPL model obtained from the above analysis, with its spectral-shape
parameters fixed. The normalizations of LS~5039 and of neighboring sources within $5^\circ$ of
the ROI center were left free, while the parameters of all other sources were fixed at their
best-fit values.

The resulting SED is shown in Figure~\ref{fig:allsed}. At lower energies, our measurements
are broadly consistent with those reported in previous studies (e.g., \citealt{yke+21}). However,
the SED exhibits a more complex spectral structure at higher energies.
A high-energy excess is detected in the $\sim30$--$500$~GeV relative to
the extrapolation of the lower-energy spectral trend. We quantify the significance of the
additional component by comparing the BPL and BPL+LP fits over
0.1--500.0~GeV, using identical event selections and background treatments.
For the additional LP component, the reference energy is fixed at 100~GeV,
while its normalization, spectral index, and curvature are allowed to vary freely. We
define
$
\Delta{\rm TS} =2\left[\ln\mathcal{L}_{\max}({\rm BPL+LP}) - \ln\mathcal{L}_{\max}({\rm BPL})\right].
$
Adopting an asymptotic $\chi^2_3$ approximation for the three additional
parameters\footnote{\url{https://fermi.gsfc.nasa.gov/ssc/data/analysis/likelihood/likelihood_overview.html}},
we obtain $\Delta{\rm TS}=72.34$ for the phase-averaged spectrum,
corresponding to a significance of $7.90\sigma$ with $p = 1.35\times10^{-15}$.
No additional scans or repeated testing were performed, hence the trials factor is taken to be unity.
%Assuming an effective trials factor of $N_{\rm trial}=500$ to approximately account for the look-elsewhere effect, the corresponding global significance decreases to $7.09\sigma$ ($p_{\rm global}=6.73\times10^{-13}$).

\begin{figure}
\centering
\includegraphics[angle=0,scale=0.65]{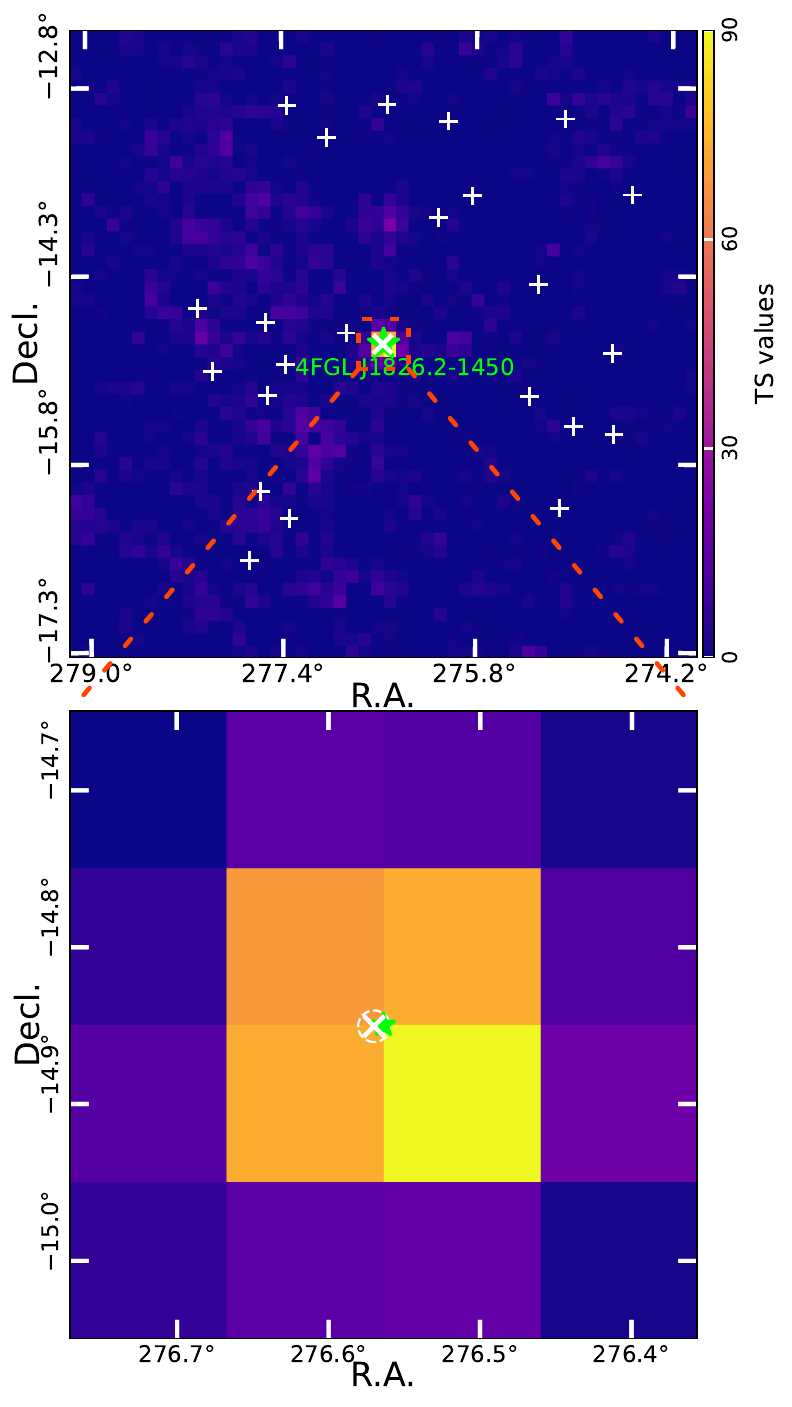}
\caption{
TS map of the $>30$~GeV $\gamma$-ray emission from LS~5039.
The upper panel shows a $5^\circ\times5^\circ$ region with
$0.1^\circ\times0.1^\circ$ pixels, centered on
4FGL~J1826.2$-$1450, with the target source excluded from
the best-fit model.
The green star marks the catalog position of
4FGL~J1826.2$-$1450, while the white crosses indicate other
$\gamma$-ray sources in 4FGL-DR4.
The lower panel shows a zoomed in view of the excess covering
a $0.4^\circ\times0.4^\circ$ region.
The white cross marks the best-fit position obtained with
\texttt{gtfindsrc}, while the green star marks the catalog
position.
The dashed white circle represents the 68\% statistical
localization region.}
\label{fig:tsmap}
\end{figure}

To investigate whether the detected high-energy excess is associated
with LS~5039, we constructed a TS map using events with energies above
30~GeV, as shown in Figure~\ref{fig:tsmap}. The significant
$\gamma$-ray excess in the TS map is spatially coincident with the
nominal position of LS~5039, with the high-energy photons concentrated
around the source location. At $E>30$~GeV,
%the Fermi-LAT point-spread function (PSF) is relatively narrow, with a 68\% containment radius below $\sim0^\circ.11$ for the relevant event selection,
%\footnote{\url{https://fermi.gsfc.nasa.gov/ssc/data/analysis/documentation/Cicerone/Cicerone_LAT_IRFs/IRF_PSF.html}} providing improved source-localization capability.
We further refined the position of the excess using
\texttt{gtfindsrc}. The best-fit position is
$\mathrm{R.A.}=276.^\circ5700$ and
$\mathrm{Decl.}=-14.^\circ8505$ (J2000), with a 68\% statistical
localization radius of $\sim0.^\circ01$. 
%Assuming a circular two-dimensional Gaussian localization, the corresponding 95\% radius is $0.0164^\circ$.
The best-fit position is separated from the position of 4FGL~J1826.2$-$1450 by $\sim0.^\circ006$, placing the catalog position within
the 68\% error circle.
Hence, we conclude that the high-energy $\gamma$-ray excess is
very likely associated with LS~5039, rather than arising from
contamination by nearby $\gamma$-ray sources or diffuse-background
emission.

%%%%%%%%%%%%
%%%%%%%%%%%%

\subsection{Phase-Resolved Analysis}
%%%%%%
\begin{figure*}
	\centering
	\includegraphics[angle=0,scale=0.55]{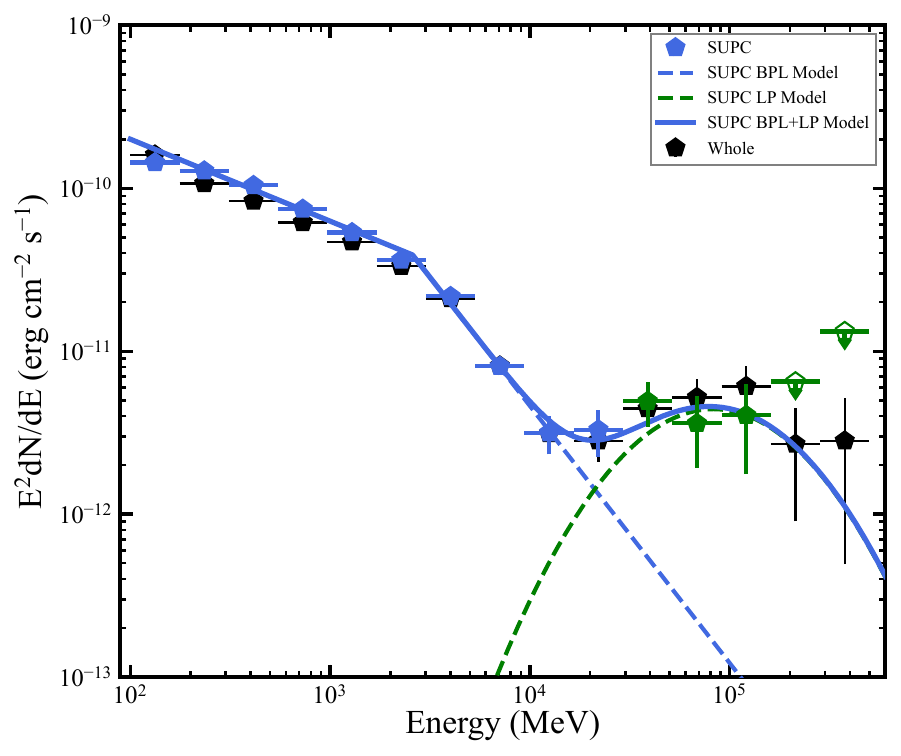}
    \includegraphics[angle=0,scale=0.55]{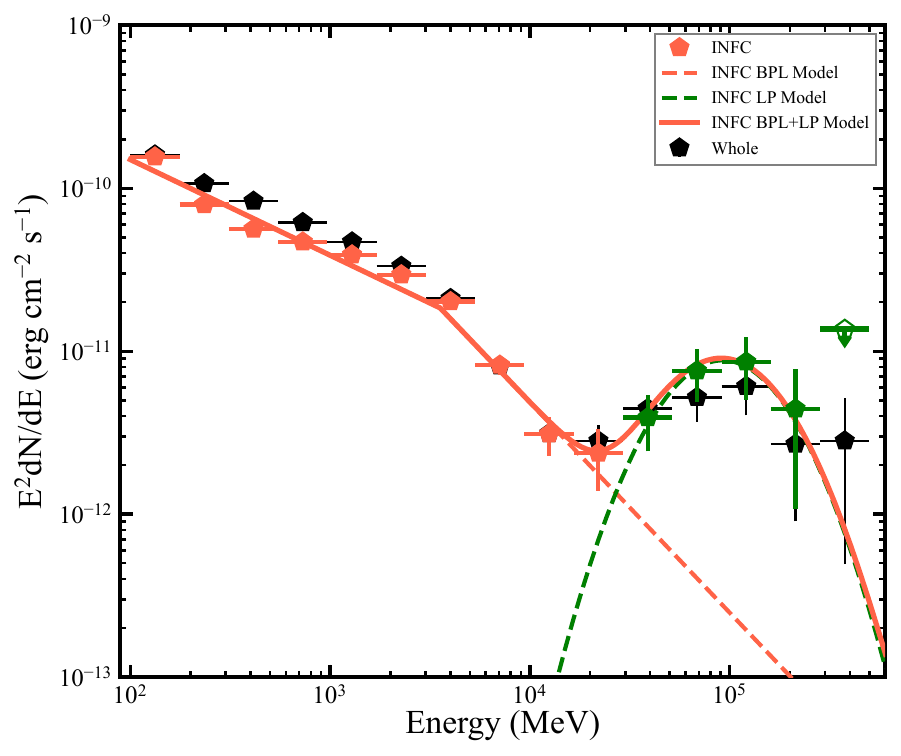}
	\caption{SEDs for the SUPC and INFC intervals, shown in blue and red
in the left and right panels, respectively. For comparison, the
phase-averaged SED from Figure~\ref{fig:allsed} is shown as black
points. In both panels, the $\gamma$-ray excess at
$\sim30$--$500$~GeV is highlighted in green. The blue and red dashed
curves represent the BPL components for SUPC and INFC, respectively,
the green dashed curves show the additional LP components, and the
corresponding solid curves show the combined BPL+LP models. Open green
symbols with downward arrows denote 95\% upper limits.}
	\label{fig:phsed}
\end{figure*}
%%%%%%
To investigate whether the $\gamma$-ray excess is modulated with the orbital phase, we constructed
phase-resolved spectra for the superior conjunction (SUPC) and inferior conjunction (INFC) intervals.
Our 17.3 yr \textit{Fermi}-LAT timing analysis gives
$P_{\rm orb}=3.90609\,{\rm d}$, consistent with \citet{zzr+24}.
We adopt phase zero at periastron,
$T_0={\rm HJD}\,2451943.09$ (${\rm MJD}\,51942.59$), and define
SUPC as $0\leq\phi<0.45$ and $0.9\leq\phi<1.0$ and INFC as
$0.45\leq\phi<0.9$, following \citet{aab+06,aaa+25}.
As noted by \citet{yke+21}, orbital-phase uncertainties can
accumulate over long observational baselines owing to uncertainties
in the orbital period and reference epoch. The published H.E.S.S.\
and HAWC phase-resolved data use the same orbital-phase zero point as
adopted here but $P_{\rm orb}=3.90603\,{\rm d}$, differing from our
value of $3.90609\,{\rm d}$ by only
$6\times10^{-5}\,{\rm d}$ ($\sim0.0015\%$). The resulting phase
offset,
$\Delta\phi=(t-T_0)(1/3.90609-1/3.90603)$, reaches at most
$|\Delta\phi|\simeq0.036$, corresponding to only $\sim6.5\%$ and
$\sim8.0\%$ of the SUPC and INFC phase widths, respectively.
Thus, this period difference does not materially affect the
phase-resolved spectral comparison, except possibly near the phase
boundaries.

For the SUPC and INFC intervals, we repeated the likelihood analysis
using the same event selection, source model, and diffuse backgrounds
as in the phase-averaged analysis, selecting only events in the
corresponding phase range. Independent broadband fits were performed
for the two intervals. The best-fit BPL+LP parameters
are listed in Table~\ref{tab:bpl_lp_parameters}, and the corresponding
models are shown in Figure~\ref{fig:phsed}.
The phase-resolved SEDs were then extracted by fixing the target
spectral shape in each energy bin to the best-fit results, while
allowing its normalization and those of sources within $5^\circ$ to
vary. Bins with $\mathrm{TS}<4$ are reported as 95\% upper limits.
The phase-averaged SED is also shown for comparison, and the INFC
spectrum exhibits a more complex structure than the SUPC spectrum.
Notably, the $\gamma$-ray excess in the $\sim$30--500~GeV (colored green) is predominantly
detected during the INFC interval. These phase-resolved SEDs are used in the following sections
to investigate the orbital modulation of the broadband $\gamma$-ray emission.

\section{Radiative Mechanisms and Particle Cooling}
\label{sec:rapa}

To model the broadband SED of LS~5039, we adopt a hybrid
leptonic--hadronic framework in which electrons and protons are
accelerated at the IBS formed by the interaction of the pulsar and
stellar winds. The model includes IC emission from relativistic
electrons \citep{kak14}, $pp$ emission from accelerated protons
\citep{kat+14}, and a phase-independent magnetospheric component
described by the outer-gap model of \citet{tlt+14}, following
\citet{wtc10}. High-energy $\gamma$-rays produced at the IBS are
attenuated by $\gamma\gamma$ absorption in the stellar radiation
field, which we account for by multiplying the intrinsic spectrum by
$\exp(-\tau_{\gamma\gamma})$; the energy- and phase-dependent optical
depth is calculated following \citet{dubus06} with the full orbital
geometry.

\begin{figure}[tbp]
	\centering
	\includegraphics[angle=0,scale=0.35]{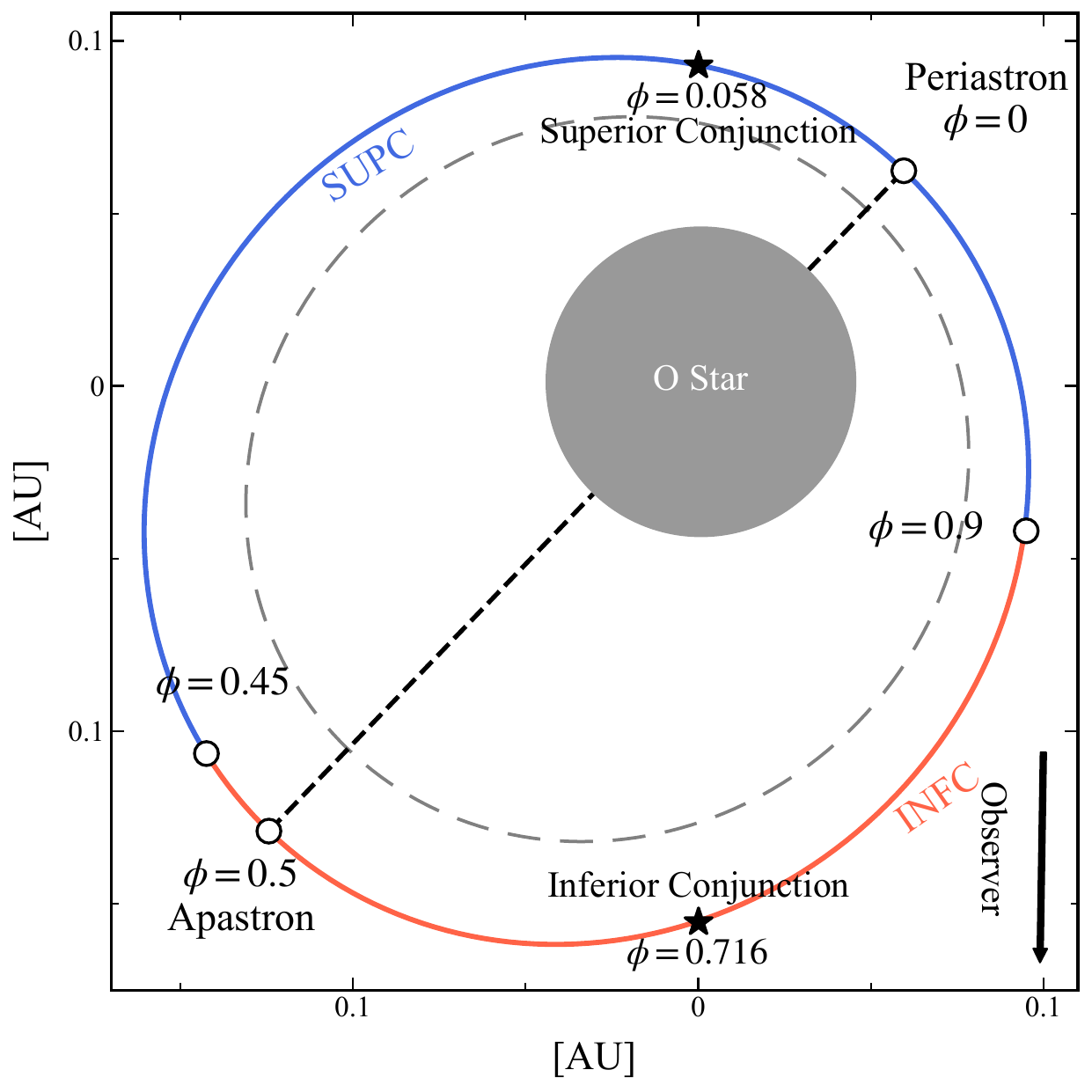}
	\caption{Schematic orbital geometry of LS~5039 based on the orbital parameters
             reported by \citet{crr+05}. The gray circle represents the O6.5\,V((f))
             companion star, while the neutron star follows an eccentric orbit with
             $e=0.35$. The blue and red segments denote the SUPC and INFC intervals,
             respectively \citep{aab+06}. The gray dashed ellipse shows the locus of
             the IBS apex over the orbit for a wind-momentum ratio of $\eta=0.05$.
             Periastron, apastron, superior conjunction, and inferior conjunction
             occur at orbital phases $\phi=0$, $0.5$, $0.058$, and $0.716$, respectively.
             The black dashed lines indicate the periastron and apastron separations,
             and the arrow denotes the direction toward the observer.
             The orbital-geometry schematic is adapted from \citet{hkr21}.}
	\label{fig:orbital}
\end{figure}

We adopt the orbital and stellar parameters of \citet{crr+05}, with
the updated period $P_{\rm orb}=3.90609$~d \citep{zzr+24},
$T_{\rm eff}=3.9\times10^{4}$~K, and
$L_\star\simeq1.8\times10^{5}L_\odot$. Motivated by the possible
hard-X-ray pulsation \citep{yme+20}, we assume a
$1.4\,M_\odot$ neutron star with $i=60^\circ$, $e=0.35$, and
$a=0.145$~AU. The corresponding periastron and apastron separations
are 0.094 and 0.196~AU, respectively. The adopted geometry and
SUPC/INFC intervals are shown in Figure~\ref{fig:orbital}.
The IBS apex is set by the wind-momentum balance,
\begin{equation}
    \eta = \frac{L_{\rm sd}}{\dot{M} v_w c},
\end{equation}
where $L_{\rm sd}$, $\dot{M}$, and $v_w$ are the pulsar spin-down
power, stellar mass-loss rate, and wind velocity
\citep{eu93,crw96}. Adopting
$L_{\rm sd}=2\times10^{36}$~erg~s$^{-1}$,
$\dot{M}=10^{-7}\,M_\odot$~yr$^{-1}$, and
$v_w=2\times10^{8}$~cm~s$^{-1}$ \citep{tlt+14} gives
$\eta\simeq0.05$. For isotropic winds,
$
R_\star(\phi)=\frac{D_{\rm sep}(\phi)}{1+\sqrt{\eta}},
R_{\rm p}(\phi)=
\frac{\sqrt{\eta}\,D_{\rm sep}(\phi)}{1+\sqrt{\eta}},
$
so that
$R_\star(\phi)\simeq0.82D_{\rm sep}(\phi)$ and
$R_{\rm p}(\phi)\simeq0.18D_{\rm sep}(\phi)$.
The apex therefore follows the phase-dependent binary separation;
we adopt this simplified local geometry throughout.

\begin{deluxetable}{lcc}
\tablecaption{Phase-weighted physical parameters for the SUPC and INFC intervals.\label{tab:fitpara}}
\tablewidth{0pt}
\tablehead{
\colhead{Parameter}
& \colhead{SUPC interval}
& \colhead{INFC interval}
}
\startdata
$\langle u_\star\rangle_\phi\;(\mathrm{erg\,cm^{-3}})$
& $7.08\times10^{2}$
& $4.29\times10^{2}$ \\
$\langle D_{\rm sep}\rangle_\phi\;(\mathrm{AU})$
& $0.141$
& $0.170$ \\
$\langle\theta_{\rm IC}\rangle_\phi\;(\mathrm{deg})$
& $103.4$
& $42.0$ \\
$\langle n_{\rm H}\rangle_\phi\;(\mathrm{cm^{-3}})$
& $6.19\times10^{8}$
& $3.75\times10^{8}$ \\
$\langle B\rangle_\phi\;(\mathrm{G})$
& $3.21$
& $2.55$ \\
$\langle E_{e,\max}\rangle_\phi\;(\mathrm{TeV})$
& $34.10$
& $37.72$ \\
\tableline
$i\;(\mathrm{deg})$
& \multicolumn{2}{c}{$60$} \\
$d\;(\mathrm{kpc})$
& \multicolumn{2}{c}{$2.5$} \\
$\epsilon_B$
& \multicolumn{2}{c}{$10^{-2}$} \\
$\eta_{\rm acc,e}$
& \multicolumn{2}{c}{$1$} \\
$\eta_{\rm acc,p}$
& \multicolumn{2}{c}{$1$} \\
$E_{p,\max}\;(\mathrm{PeV})$
& \multicolumn{2}{c}{$0.346$} \\
\enddata
\tablecomments{The SUPC interval is defined as $0\leq\phi<0.45$ and $0.9\leq\phi<1.0$, while the INFC interval is defined as $0.45\leq\phi<0.9$. The quantities $u_\star$, $D_{\rm sep}$, $\theta_{\rm IC}$, $n_{\rm H}$, $B$, $E_{e,\max}$, and $E_{p,\max}$ denote the stellar photon-field energy density, binary separation, IC scattering angle, target gas density, magnetic-field strength, maximum electron energy, and maximum proton energy, respectively. The quantities $i$, $d$, $\epsilon_B$, $\eta_{\rm acc,e}$, and $\eta_{\rm acc,p}$ denote the orbital inclination, source distance, magnetic-energy fraction, electron acceleration parameter, and proton acceleration parameter, respectively. Angle brackets denote normalized orbital-phase-weighted averages over the corresponding phase interval.}
\end{deluxetable}

The stellar photon energy density at the shock apex is
\begin{equation}
	u_\star(\phi)=\frac{L_\star}{4\pi R_\star(\phi)^2c},
\end{equation}
while the proton number density of the stellar wind is
\begin{equation}
	n_{\rm H}(\phi)=\frac{\dot{M}}{4\pi R_\star(\phi)^2v_wm_p},
\end{equation}
where $m_p$ is the proton mass. These phase-dependent quantities serve as the target photon field for IC scattering and the target matter field for $pp$ interactions in the SED modeling.

We assume that the IBS magnetic field originates in the pulsar wind
and parameterize the post-shock magnetic energy density as
$U_B(\phi)=B(\phi)^2/8\pi$, with $\epsilon_B$ denoting the magnetic
fraction of the local pulsar-wind energy density:
\begin{equation}
    \frac{B(\phi)^2}{8\pi} = \epsilon_B\frac{L_{\rm sd}}{4\pi R_{\rm p}(\phi)^2c}.
\end{equation}
Neglecting shock compression and field geometry,
$\epsilon_B\sim\sigma/(1+\sigma)\simeq\sigma$ for $\sigma\ll1$.
We adopt $\epsilon_B=10^{-2}$, representative of a weakly magnetized
pulsar wind and comparable to values used for shocked pulsar winds
and $\gamma$-ray binaries
\citep{kc84,dubus13,tlt+14,bosch-ramon21}.

Following the orbital geometry shown in Figure~\ref{fig:orbital}, we
convert the orbital phase $\phi$ into the true anomaly $f(\phi)$ and
compute the instantaneous binary separation $D_{\rm sep}(\phi)$ from
the Keplerian orbit. The SUPC and INFC intervals are defined according
to Figure~\ref{fig:orbital}. The SUPC and INFC SEDs are integrated over their full phase
intervals using normalized Gauss--Legendre quadrature with uniform
phase weighting. We use 24 SUPC nodes, split at the occultation
boundaries, and 6 INFC nodes. Orbital geometry and local IBS
conditions are evaluated at each node; Table~\ref{tab:fitpara}
lists the corresponding phase-weighted averages.

As listed in Table~\ref{tab:fitpara}, the stellar photon energy density
at the IBS exceeds the magnetic energy density
($u_\star\gg U_B$), so IC cooling is efficient in the Thomson regime
\citep{bg70,rl79,dubus06,bk09}. For
$T_{\rm eff}\simeq3.9\times10^{4}$~K, the characteristic stellar
photon energy is $\epsilon_\star\simeq2.8kT_{\rm eff}\simeq9$~eV.
The Thomson--Klein--Nishina transition,
$4\gamma\epsilon_\star\gtrsim m_ec^2$, occurs at
$
E_{e,\rm KN}\simeq
\frac{m_e^2c^4}{4\epsilon_\star}
\simeq7~{\rm GeV}
$
\citep{bg70,msc+05,kak14}. Above this energy, KN suppression reduces
the IC cooling rate \citep{msc+05,kab08,kak14}, allowing synchrotron
losses to determine the maximum electron energy despite
$u_\star\gg U_B$ \citep{msc+05,dubus06,bk09,lcc+25}. With
$\eta_{\rm acc}\geq1$ parameterizing the acceleration efficiency
($\eta_{\rm acc}=1$ at the Bohm limit), balancing acceleration and
synchrotron losses gives \citep{kab08,dlf15}
\begin{equation}
	E_{e,\max}(\phi)
	\simeq
	60\,
	\eta_{\rm acc}^{-1/2}
	\left[
	\frac{B(\phi)}{1~{\rm G}}
	\right]^{-1/2}
	{\rm TeV}.
\end{equation}
For $\epsilon_B=10^{-2}$ and $\eta_{\rm acc}=1$, the node-wise
fields are $2.18$--$4.48$~G in SUPC and $2.16$--$3.56$~G in INFC,
giving $E_{e,\max}=28.16$--$40.35$ and $31.58$--$40.54$~TeV.
Since $E_{e,\max}^{\rm syn}\propto
\eta_{\rm acc}^{-1/2}\epsilon_B^{-1/4}$,
$\eta_{\rm acc}=3$ gives $\sim16$--$23$~TeV, whereas
$\epsilon_B=10^{-4}$ gives a formal $\sim89$--$128$~TeV limit but
reduces the local confinement limit from $\sim0.346$ to
$\sim0.035$~PeV. Thus, lowering $B$ alone cannot both accelerate and
confine $\sim120$~TeV electrons at the IBS apex.
The phase-resolved H.E.S.S.\ and HAWC spectra show TeV orbital
modulation, with HAWC extending beyond $\sim100$~TeV
\citep{aab+06,aaa+25}. Since KN-regime IC photons satisfy
$E_\gamma\lesssim E_e$, the $\sim30$--$40$~TeV electron limits
cannot reproduce the $>100$~TeV emission in the local IBS model.
We therefore test a phenomenological pion-decay component from
$pp$ interactions \citep{kab06}.

A possible origin of the $>100$~TeV emission is pion decay from
inelastic $pp$ interactions of shock-accelerated protons with the
stellar wind \citep{kab06}. The effective $pp$ efficiency depends on
the proton residence time and mixing across the contact discontinuity,
which are not resolved in our one-zone treatment. We therefore include
the hadronic component phenomenologically, with the maximum proton
energy constrained by acceleration, cooling, escape, and confinement
in the local IBS environment.

Equating $t_{\rm acc}=\eta_{\rm acc,p}E_p/(eBc)$ with
$t_{pp}=(\kappa_{pp}\sigma_{pp}n_{\rm H}c)^{-1}$ gives
$E_{p,\max}^{(pp)}\sim10^{19}$--$10^{20}$~eV, showing that
$pp$ cooling does not limit the proton energy under the adopted IBS
conditions. A more restrictive local estimate follows from the Hillas confinement
condition, $E_{p,\max}^{\rm conf}\simeq eBR_{\rm acc}$, where
$R_{\rm acc}$ is the characteristic acceleration and confinement
scale. Identifying this scale with the pulsar-to-apex distance,
$R_{\rm acc}=R_{\rm p}$, gives
$E_{p,\max}^{\rm conf}\sim3\times10^{14}$~eV.
We adopt
$E_{p,\max}=\min(E_{p,\max}^{(pp)},E_{p,\max}^{\rm Hillas},
E_{p,\max}^{\rm dyn})$, with
$E_{p,\max}^{\rm Hillas}=eBR_{\rm p}$ and
$E_{p,\max}^{\rm dyn}=eBR_{\rm p}/\eta_{\rm acc,p}$.
For $\eta_{\rm acc,p}=1$,
$BR_{\rm p}=(2\epsilon_B L_{\rm sd}/c)^{1/2}$ gives the local limit
$E_{p,\max}\simeq0.346$~PeV. Since
$E_{p,\max}^{\rm dyn}\propto
\epsilon_B^{1/2}\eta_{\rm acc,p}^{-1}$,
$\eta_{\rm acc,p}=3$ gives $\sim0.115$~PeV and
$\epsilon_B=10^{-4}$ gives $\sim0.035$~PeV. Larger acceleration
scales are not constrained by the present local model.
The maximum particle energies are imposed in \textsc{Naima}
\citep{zabalza15} as upper integration limits rather than explicit
cutoffs in the analytic particle distributions. The IC and pion-decay
calculations are therefore restricted to
$E_e\leq E_{e,\max}(\phi)$ and $E_p\leq E_{p,\max}$, respectively.

\section{Spectral Fitting Procedure}
\label{sec:specfit}

Based on the hybrid leptonic--hadronic radiation model and the binary geometry described in
Section~\ref{sec:rapa}, we perform independent broadband spectral fits to the GeV--UHE SEDs
of LS~5039 for the SUPC and INFC intervals using the corresponding phase-resolved data sets.
The calculations are carried out with the Python package \textsc{Naima} (v0.10; \citealt{zabalza15}),
including IC emission and $pp$ interactions. Within each conjunction interval, particle-distribution
parameters are shared among all phase nodes, while
$u_\star$, $B$, $n_{\rm H}$, the IC angle, particle-energy limits,
and $\gamma\gamma$ opacity are evaluated locally. The phase-averaged
SED is
\begin{equation}
\begin{aligned}
\overline{F}_{r}(E) ={}& F_{\rm OG}(E)
+\sum_{i\in r} w_i
\Bigl[
F_{\rm IC1}(E,\phi_i)
+F_{\rm IC2}(E,\phi_i) \\
&\qquad\qquad
+F_{pp}(E,\phi_i)
\Bigr]
\exp[-\tau_{\gamma\gamma}(E,\phi_i)] .
\end{aligned}
\end{equation}
Here $\sum_iw_i=1$; $F_{\rm OG}$ is phase independent and unattenuated,
and $F_{pp}=0$ in the purely leptonic model. This treatment ensures consistency with the varying
binary environment implied by the eccentric orbital geometry and reduces degeneracies between
these environmental quantities and the parameters describing the particle distributions,
including their normalizations, characteristic energies, spectral breaks,
and cooling-related quantities. We fit the detected GeV--UHE flux points with the Gaussian
likelihood implemented in \texttt{naima},
\begin{equation}
\ln\mathcal{L}=
-\frac{1}{2}\sum_{i=1}^{N}
\left[
\frac{F_i-F_{\rm mod}(E_i;\boldsymbol{\theta})}{\sigma_i}
\right]^2 ,
\end{equation}
where $\sigma_i$ denotes the adopted flux uncertainty. The posterior
distributions are explored using an ensemble Markov Chain Monte Carlo
(MCMC) sampler. We use $N=27$ SUPC and 46 INFC detections; upper
limits and X-ray data are excluded from the likelihood and BIC. No
additional cross-instrument calibration uncertainties or correlated
covariance terms are included.

The model components dominate different, although partially
overlapping, parts of the SED. The phase-independent magnetospheric
component and IC1 from the relativistic-Maxwellian electron population
mainly describe the GeV emission, while IC2 from the broken-power-law
electrons accounts for the TeV component. The hadronic $pp$ component
contributes mainly to the UHE tail detected by HAWC. X-ray data are
not included in the present fit, as this study focuses on the
$\gamma$-ray and higher-energy emission of LS~5039, spanning the GeV
to UHE regime. This energy range provides the most direct constraints
on the relative contributions of the leptonic and hadronic emission
channels.

\begin{figure*}
	\centering
	\includegraphics[angle=0,scale=0.405]{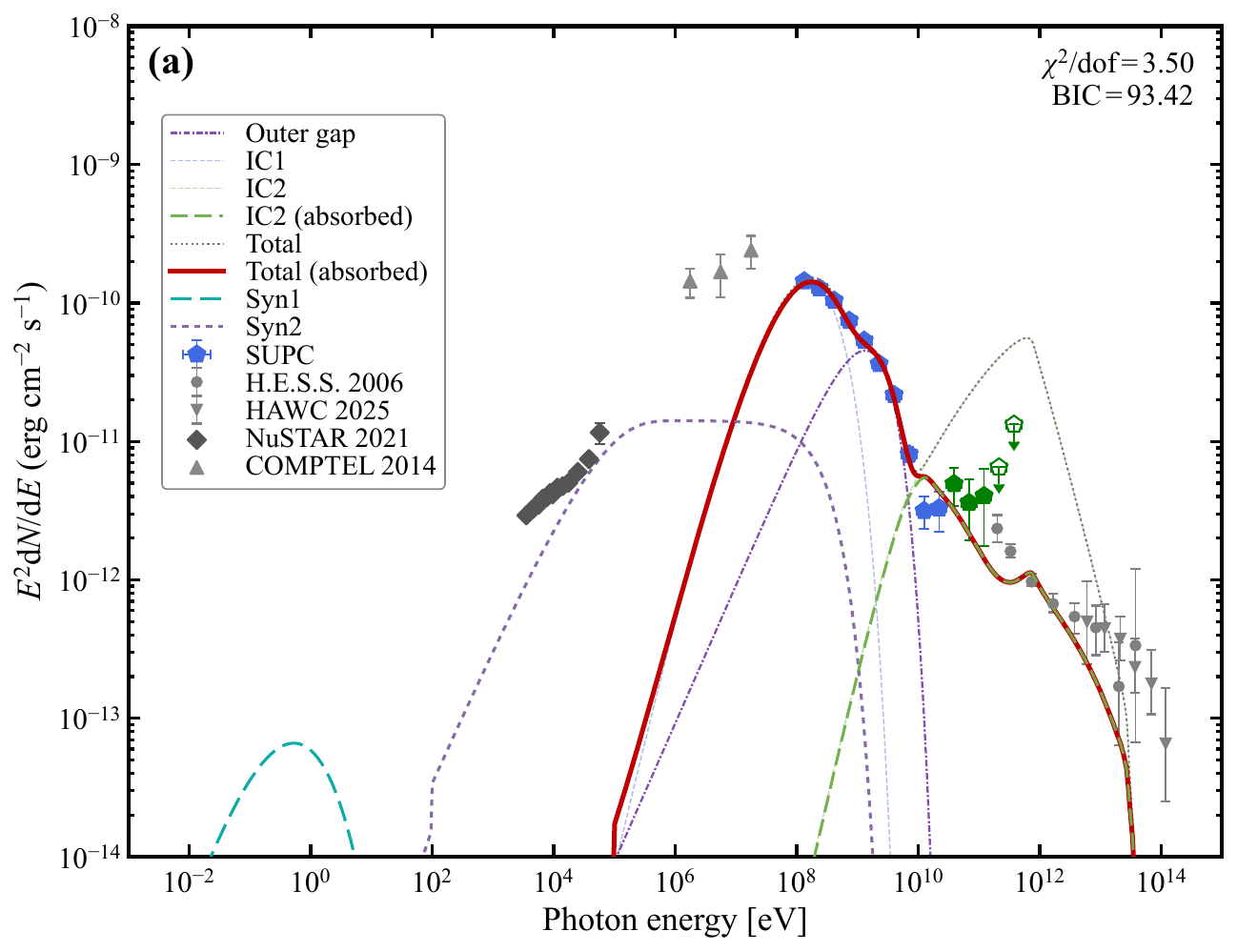}
    \includegraphics[angle=0,scale=0.405]{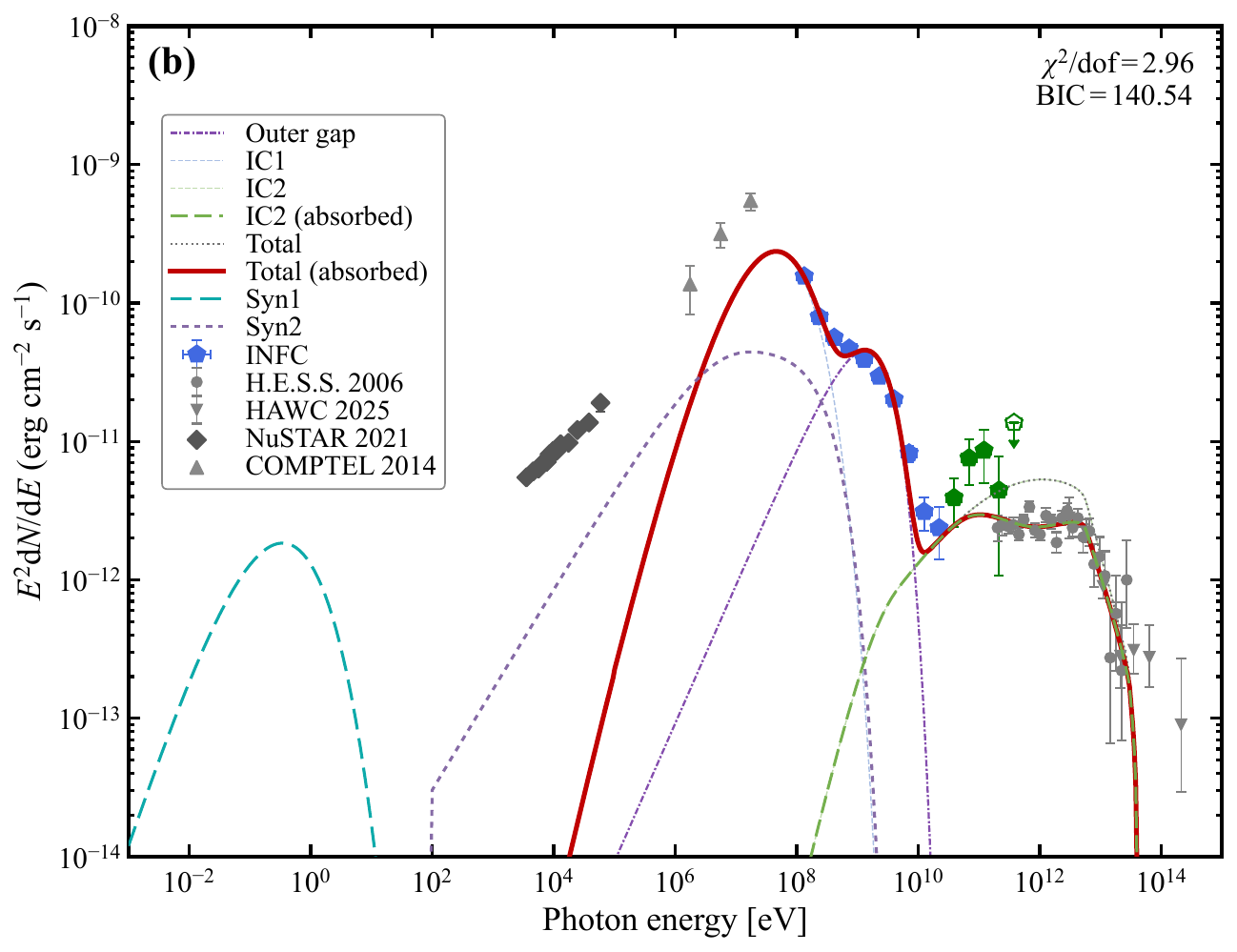}
    \includegraphics[angle=0,scale=0.405]{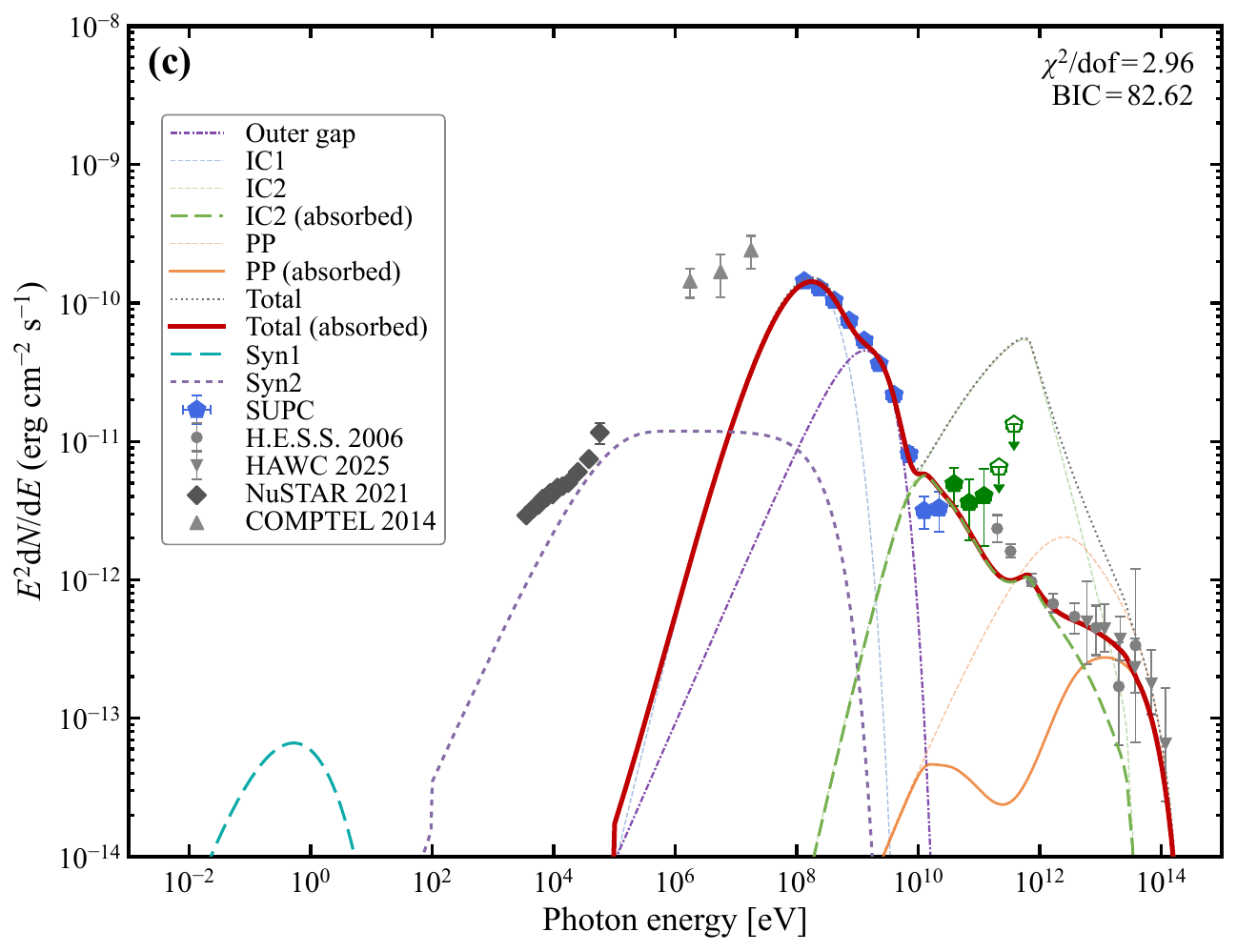}
    \includegraphics[angle=0,scale=0.405]{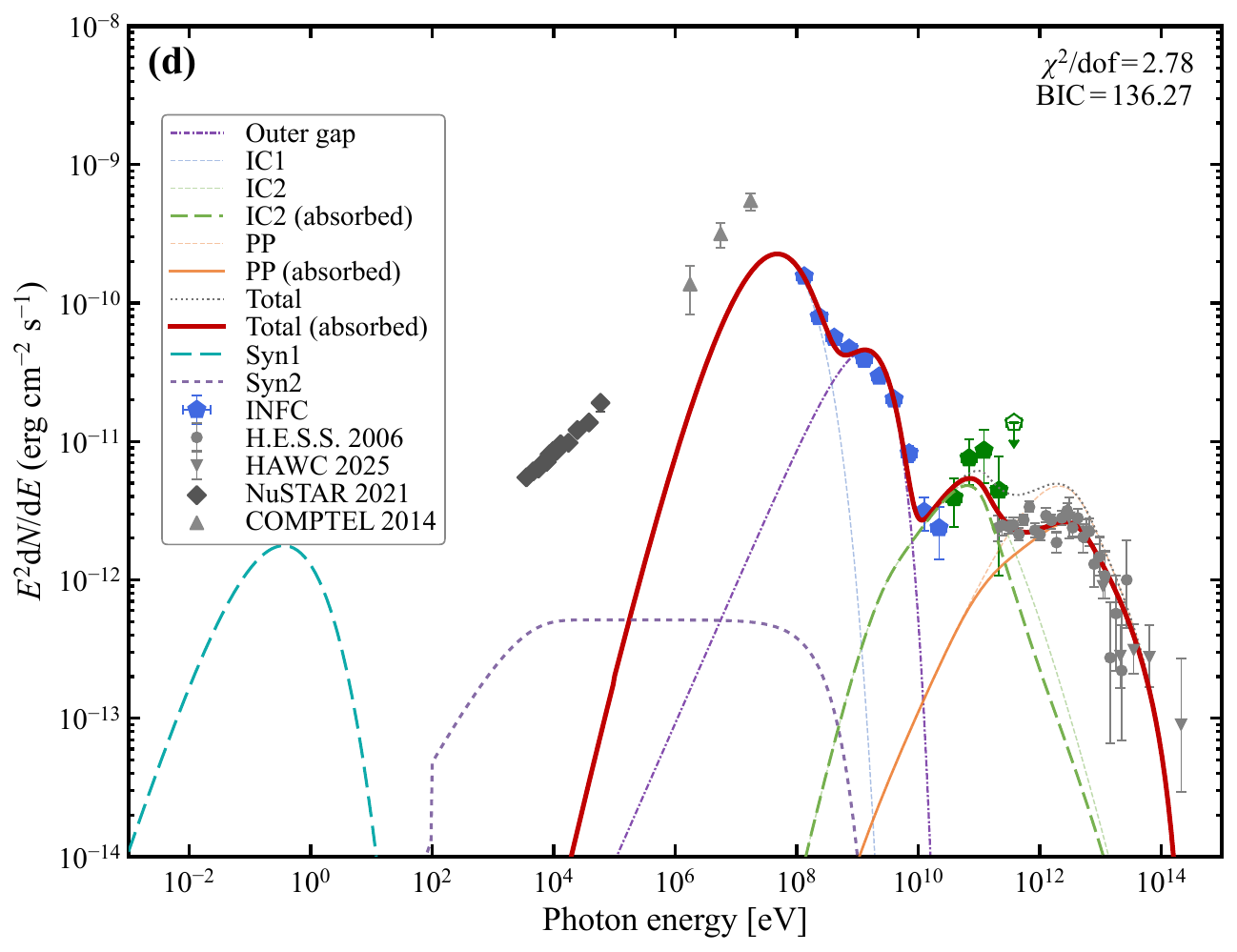}
\caption{Broadband SED fits for SUPC (left) and INFC (right):
purely leptonic models are shown in panels (a,b) and lepto-hadronic
models in panels (c,d). Purple dash-dotted, light-blue dashed, and
green curves denote the outer-gap, IC1, and IC2 components,
respectively; dark-green curves show absorbed IC2. Cyan and violet
curves show Syn1 and Syn2. In panels (c,d), pale-orange and orange
curves denote intrinsic and absorbed $pp$ emission. Gray dotted and
red solid curves show the intrinsic and absorbed total models.
Blue/green pentagons are LAT measurements below/above 30~GeV; open
symbols denote 95\% upper limits. Gray data are from H.E.S.S.\
\citep{aab+06}, HAWC \citep{aaa+25}, \textit{NuSTAR}
\citep{yke+21}, and COMPTEL \citep{cz14}; X-ray and MeV points are
not fitted.}
	\label{fig:sedfit}
\end{figure*}

\begin{figure*}
	\centering
	\includegraphics[angle=0,scale=0.425]{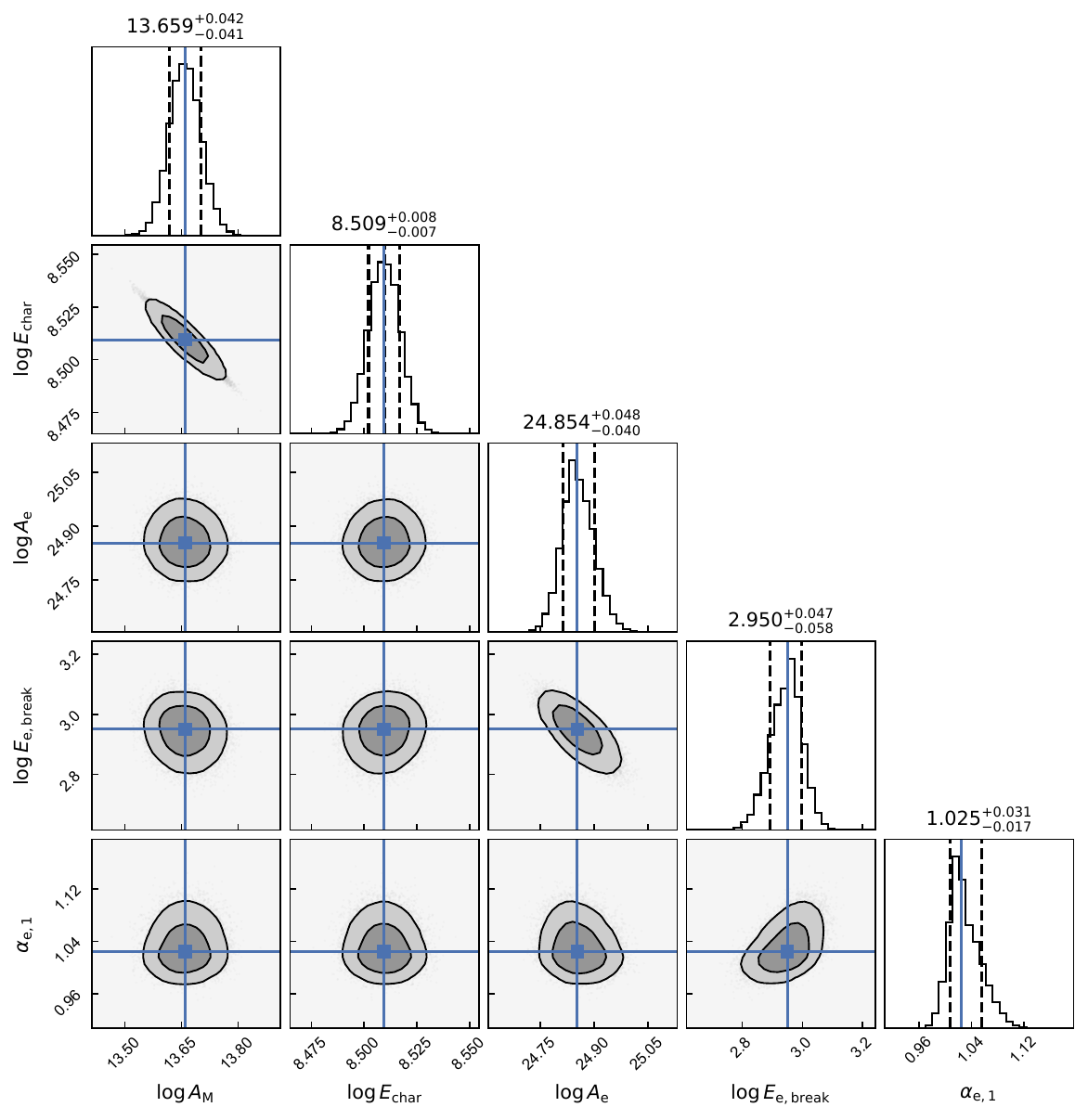}
    \includegraphics[angle=0,scale=0.425]{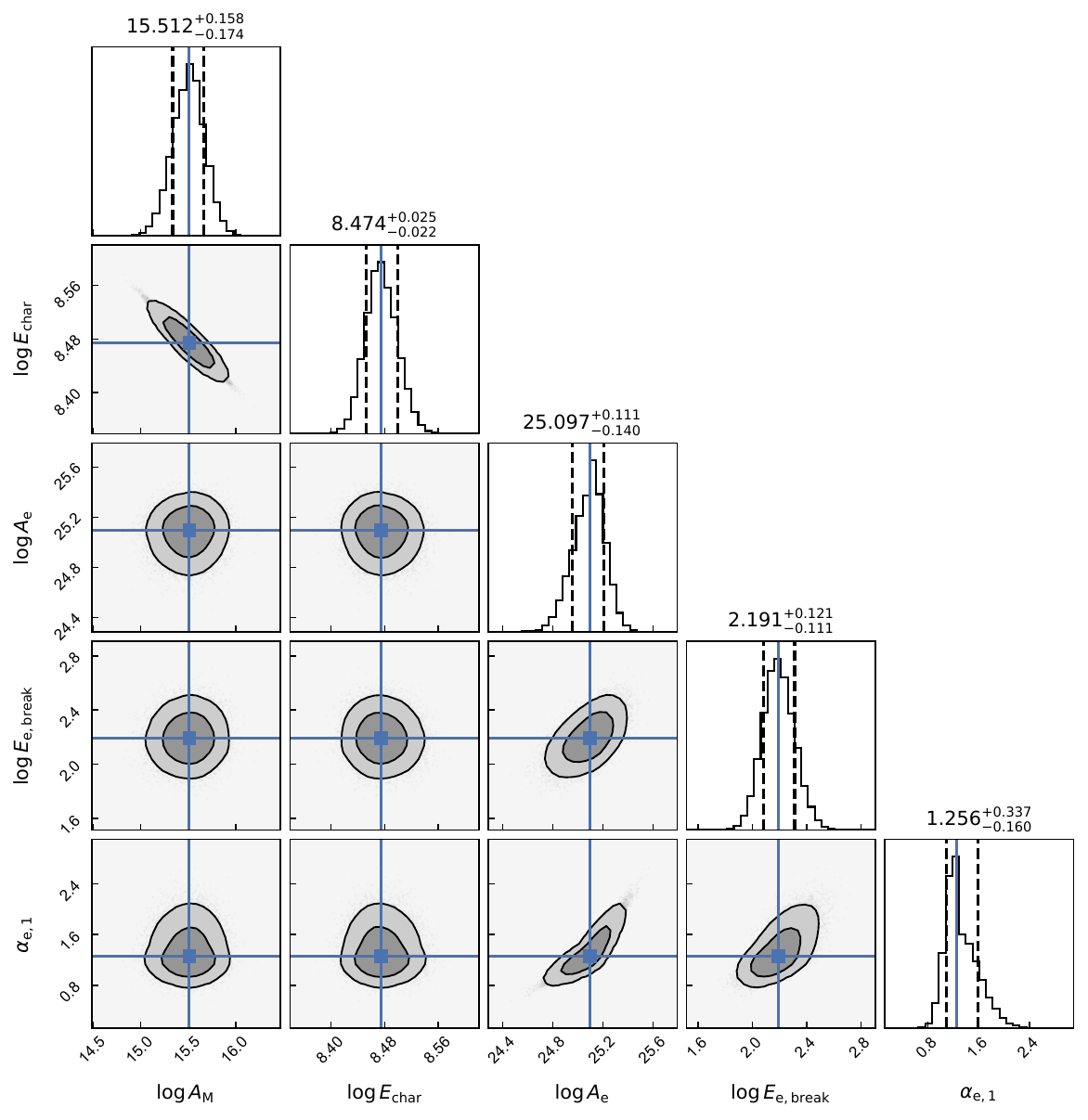}
    \includegraphics[angle=0,scale=0.32]{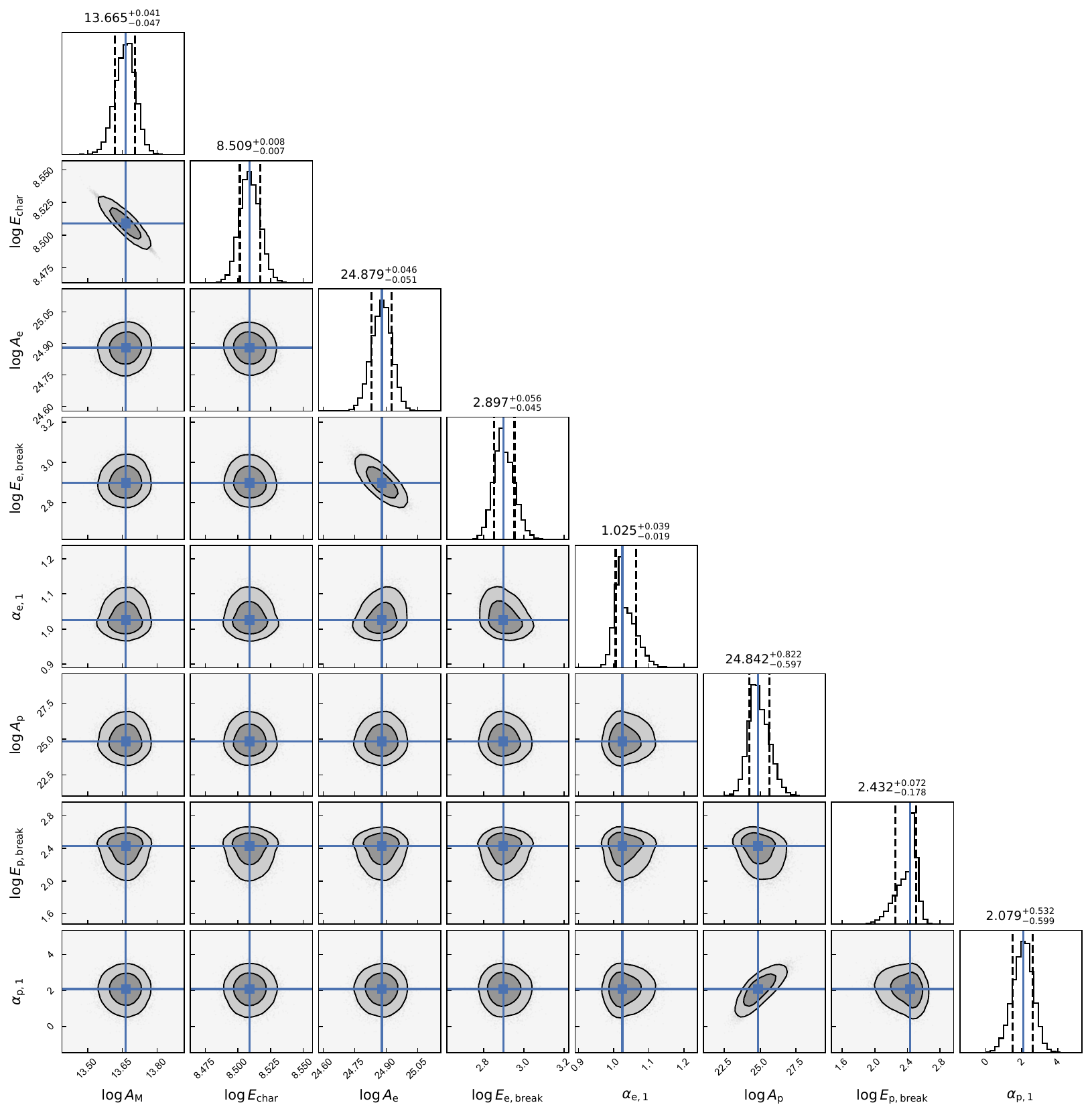}
    \includegraphics[angle=0,scale=0.32]{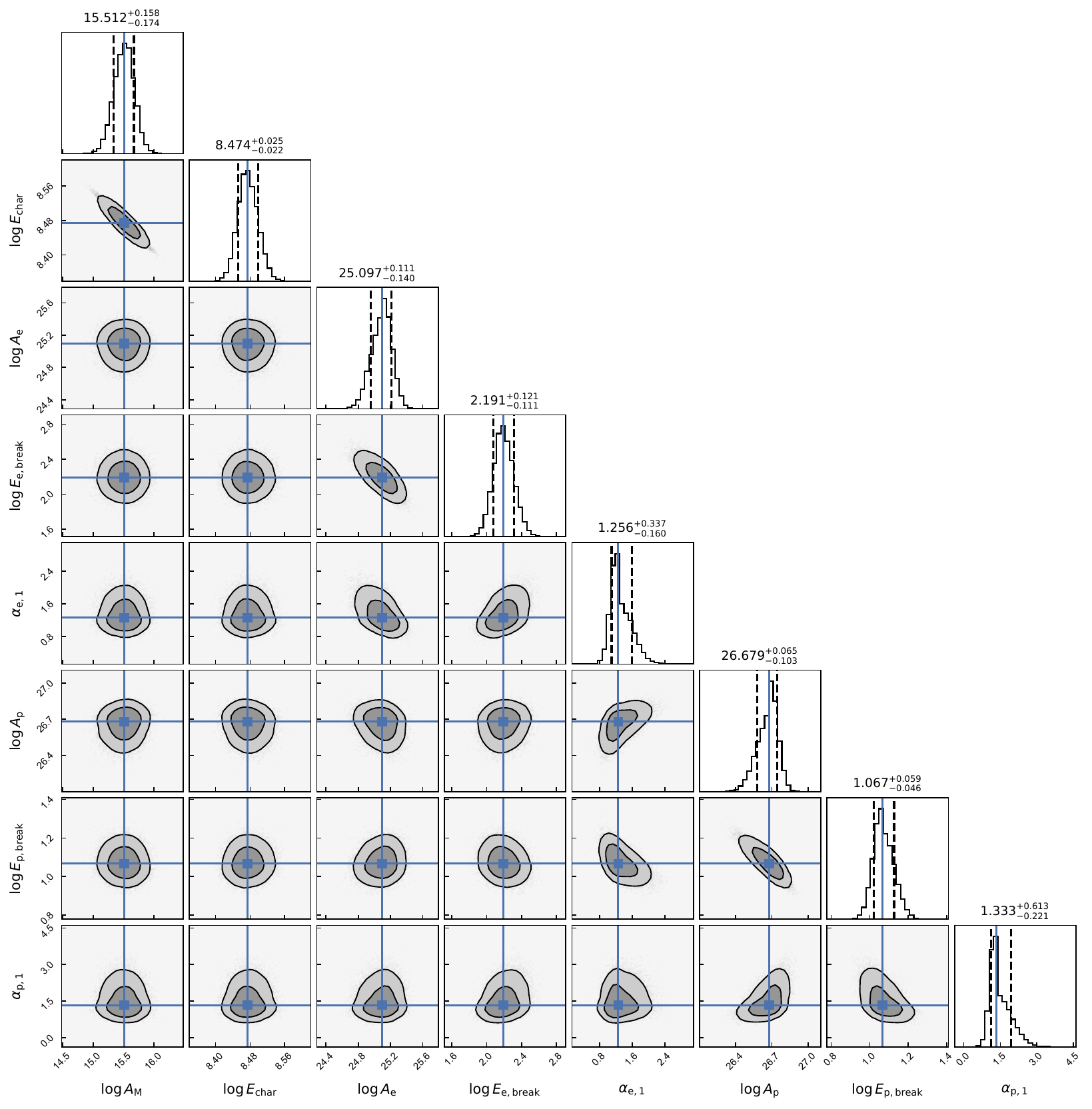}
	
	\caption{Posterior distributions for the SUPC (left) and INFC
(right) fits. The upper and lower rows correspond to the purely
leptonic and lepto-hadronic models, respectively.}
	\label{fig:corner}
\end{figure*}

In the \textsc{Naima} fitting, we consider two electron populations and one proton population.
The first electron population, responsible for the thermal electron component (corresponding to IC1),
is described by a relativistic Maxwellian distribution,
\begin{equation}
    \frac{dN_{e,{\rm M}}}{dE}
    = A_{\rm M} E^2
    \exp\left(-\frac{E}{E_{\rm char}}\right),
\end{equation}
where $A_{\rm M}$ is the normalization and $E_{\rm char}$ is the characteristic electron energy.
The second electron population, representing shock-accelerated electrons and producing the IC2
component, is described by a broken power-law distribution,
\begin{equation}
\frac{dN_{e,{\rm BKN}}}{dE}
= A_e
\begin{cases}
\left(E/E_{0,e}\right)^{-\alpha_{e,1}},
& E \leq E_{e,\mathrm{break}},
\\[2mm]
\begin{aligned}
&\left(E_{e,\mathrm{break}}/E_{0,e}\right)^{
  \alpha_{e,2}-\alpha_{e,1}}
\\[-1mm]
&\qquad {}\times
\left(E/E_{0,e}\right)^{-\alpha_{e,2}},
\end{aligned}
& E > E_{e,\mathrm{break}},
\end{cases}
\end{equation}
where $A_e$ is the normalization, $E_{e,\mathrm{break}}$ is the break energy, and $\alpha_{e,1}$ and
$\alpha_{e,2}$ are the spectral indices below and above the break energy, respectively.
We fix the reference energy and high-energy spectral index at $E_{0,e}=100$~GeV and
$\alpha_{e,2}=3$, respectively.
The proton population, which produces the hadronic emission through inelastic $pp$ interactions, is also modeled with a broken power-law distribution,
\begin{equation}
\frac{dN_{p,{\rm BKN}}}{dE}
= A_p
\begin{cases}
\left(E/E_{0,p}\right)^{-\alpha_{p,1}},
& E \leq E_{p,\mathrm{break}},
\\[2mm]
\begin{aligned}
&\left(E_{p,\mathrm{break}}/E_{0,p}\right)^{
  \alpha_{p,2}-\alpha_{p,1}}
\\[-1mm]
&\qquad {}\times
\left(E/E_{0,p}\right)^{-\alpha_{p,2}},
\end{aligned}
& E > E_{p,\mathrm{break}},
\end{cases}
\end{equation}
where $A_p$ is the proton normalization, $E_{p,\mathrm{break}}$ is the break energy, and $\alpha_{p,1}$
and $\alpha_{p,2}$ are the spectral indices below and above the break energy, respectively.
We fix $E_{0,p}=10$~TeV and $\alpha_{p,2}=3$,
and impose the physically derived $E_{p,\max}(\phi)$ as the
upper proton integration limit at each phase node. Therefore, for each phase interval,
the lepto-hadronic model has eight free parameters: $A_{\rm M},\ E_{\rm char},\ A_e,\ E_{e,\mathrm{break}},\
    \alpha_{e,1},\ A_p,\ E_{p,\mathrm{break}},\ \alpha_{p,1}$.
SUPC and INFC are fitted independently using the bounded uniform
priors in Table~\ref{tab:prior_bounds}, with 128 walkers, 1500
burn-in steps, and 10000 production steps. Mean acceptance fractions
are 0.460--0.552, maximum $\tau_{\rm int}=54.9$--$86.0$ steps,
all chains span at least $116\tau_{\rm int}$, and the minimum ESS is
$1.49\times10^{4}$--$2.33\times10^{4}$.
The posterior parameter estimates are listed in Table~\ref{tab:result}.
The corresponding broadband SED fits are shown in Figure~\ref{fig:sedfit}, and the posterior distributions are presented as corner plots in Figure~\ref{fig:corner}.

\begin{deluxetable*}{lcc}
\tablecaption{Uniform prior ranges adopted in the MCMC fits\label{tab:prior_bounds}}
\tablewidth{0pt}
\tablehead{
\colhead{Parameter}
& \colhead{Prior range}
}
\startdata
$\log_{10}[A_{\rm M}/(\mathrm{eV}^{-3})]$
& $(12,\,17)$\\
$\log_{10}(E_{\rm char}/\mathrm{eV})$
& $(7,\,10)$\\
$\log_{10}[A_e/(\mathrm{eV}^{-1})]$
& $(15,\,30)$\\
$\log_{10}(E_{e,\mathrm{break}}/\mathrm{GeV})$
& $(-3.29,\,4.45)$ \\
$\alpha_{e,1}$
& $(1,\,3)$\\
$\log_{10}[A_p/(\mathrm{eV}^{-1})]$
& $(15,\,30)$\\
$\log_{10}(E_{p,\mathrm{break}}/\mathrm{TeV})$
& $(1.00,\,2.54)$\\
$\alpha_{p,1}$
& $(1,\,3)$\\
\enddata
\tablecomments{
All priors are uniform in the sampled variables and identical for
SUPC and INFC. Electrons span
$511$~keV$\leq E_e\leq E_{e,\max}(\phi)$; because
$E_{e,\mathrm{break}}$ is shared among phase nodes, its upper prior
uses $\min[E_{e,\max}(\phi)]=28.156$~TeV. Protons span
$10$~TeV$\leq E_p\leq E_{p,\max}(\phi)$, with the break-energy upper
bound set by $E_{p,\max}=0.346$~PeV. The high-energy indices are fixed
at $\alpha_{e,2}=\alpha_{p,2}=3$.
}
\end{deluxetable*}

\begin{deluxetable*}{lcccc}
\tablecaption{Best-fit parameters and convergence diagnostics from the MCMC analysis}
\tablewidth{0pt}
\tabletypesize{\scriptsize}
\tablehead{
\multicolumn{1}{c}{Parameter}
& \multicolumn{2}{c}{SUPC}
& \multicolumn{2}{c}{INFC} \\
\cline{2-3}\cline{4-5}
& \colhead{Leptonic}
& \colhead{Lepto-hadronic}
& \colhead{Leptonic}
& \colhead{Lepto-hadronic}
}
\startdata
$\log_{10}[A_{\rm M}/(\mathrm{eV}^{-3})]$ & $13.66_{-0.04}^{+0.04}$ & $13.67_{-0.05}^{+0.04}$ & $15.44_{-0.14}^{+0.16}$ & $15.51_{-0.17}^{+0.16}$ \\
$\log_{10}(E_{\rm char}/\mathrm{eV})$
& $8.51_{-0.01}^{+0.01}$
& $8.51_{-0.01}^{+0.01}$
& $8.48_{-0.02}^{+0.02}$
& $8.47_{-0.02}^{+0.02}$ \\
$\log_{10}[A_e/(\mathrm{eV}^{-1})]$
& $24.85_{-0.04}^{+0.05}$
& $24.88_{-0.05}^{+0.05}$
& $24.84_{-0.06}^{+0.05}$
& $25.10_{-0.14}^{+0.11}$ \\
$\log_{10}(E_{e,\mathrm{break}}/\mathrm{GeV})$
& $2.95_{-0.06}^{+0.05}$
& $2.90_{-0.04}^{+0.06}$
& $3.76_{-0.05}^{+0.05}$
& $2.19_{-0.11}^{+0.12}$ \\
$\alpha_{e,1}$
& $1.02_{-0.02}^{+0.03}$
& $1.02_{-0.02}^{+0.04}$
& $1.54_{-0.05}^{+0.05}$
& $1.26_{-0.16}^{+0.34}$ \\
$\log_{10}[A_p/(\mathrm{eV}^{-1})]$
& --
& $24.84_{-0.60}^{+0.82}$
& --
& $26.68_{-0.10}^{+0.06}$ \\
$\log_{10}(E_{p,\mathrm{break}}/\mathrm{TeV})$
& --
& $2.43_{-0.18}^{+0.07}$
& --
& $1.07_{-0.05}^{+0.06}$ \\
$\alpha_{p,1}$
& --
& $2.08_{-0.60}^{+0.53}$
& --
& $1.33_{-0.22}^{+0.61}$ \\
\tableline
$\chi^2/\mathrm{d.o.f.}$
& $3.50$
& $2.96$
& $2.96$
& $2.78$ \\
$\mathrm{BIC}$
& $93.42$
& $82.62$
& $140.54$
& $136.27$ \\
\tableline
Mean acceptance fraction
& $0.5516$
& $0.4601$
& $0.5518$
& $0.4607$ \\
Maximum $\tau_{\rm int}$ (steps)
& $56.85$
& $85.95$
& $55.27$
& $84.60$ \\
Minimum $N/\tau_{\rm int}$
& $175.89$
& $116.34$
& $180.93$
& $118.21$ \\
\enddata
\tablecomments{
Best-fit parameters and MCMC diagnostics for the SUPC and INFC spectra
under the purely leptonic and lepto-hadronic models. Parameter values
are posterior medians with uncertainties given by the 16th and 84th
percentiles. Here, $\chi^2/\mathrm{d.o.f.}$ is the reduced chi-square,
BIC is the Bayesian information criterion, $\tau_{\rm int}$ is the
integrated autocorrelation time, and $N$ is the number of retained
production steps per walker. The reported maximum $\tau_{\rm int}$
and minimum $N/\tau_{\rm int}$ correspond to the least favorable
sampled parameter in each fit. All four fits have
$N/\tau_{\rm int}>100$, indicating adequate posterior sampling.
Upper limits and X-ray measurements are excluded from the likelihood
and BIC calculations.
}
\label{tab:result}
\end{deluxetable*}

We compare the purely leptonic and lepto-hadronic models using
the Bayesian information criterion (BIC; \citealt{Schwarz78};
\citealt{fkb+22}),
\begin{equation}
    {\rm BIC} = k\ln N - 2\ln \mathcal{L}_{\rm max},
\end{equation}
where $k$, $N$, and $\mathcal{L}_{\rm max}$ are the number of
free parameters, number of fitted detections, and maximum likelihood,
respectively. Lower BIC indicates relative support for a model; we
define
\begin{equation}
    \Delta{\rm BIC}
    = {\rm BIC}_{\rm lep} - {\rm BIC}_{\rm lep\text{-}had}.
\end{equation}
So that $\Delta{\rm BIC}>0$ favors the lepto-hadronic model.
BIC is evaluated independently for SUPC and INFC.

\section{Results and Discussion}
\label{sec:result}

\subsection{Purely Leptonic Fit}

Figure~\ref{fig:sedfit}(a) and (b) show the best-fit purely leptonic
SED models for the SUPC and INFC intervals. In both phases, the
emission below $\sim10$~GeV is dominated by the relativistic-Maxwellian
and magnetospheric components, while shock-accelerated electrons
connect the GeV and TeV bands. The Maxwellian parameters
$A_{\rm M}$ and $E_{\rm char}$ are similar in the two intervals,
whereas the broken-power-law electron population is phase dependent:
$(\log E_{e,\mathrm{break}}/{\rm GeV},\alpha_{e,1})=(2.95,1.02)$
for SUPC and $(3.76,1.54)$ for INFC. Thus, the INFC TeV emission
requires a higher break energy and a softer low-energy electron
spectrum than in SUPC.
%The residuals indicate that the model describes most of the GeV--TeV data reasonably well,
%although deviations remain at the highest energies.

These results are broadly consistent with previous leptonic interpretations of LS~5039, in which
anisotropic IC scattering, $\gamma\gamma$ absorption, and orbital-phase-dependent radiative
cooling play important roles in shaping the observed GeV--TeV spectrum \citep{dch08,kab08,zba+13}.
Compared with simple one-zone IC models, our phase-resolved fits suggest that
a single homogeneous leptonic emitter may not fully account for all of the observed spectral
features, particularly when the highest-energy TeV measurements are included \citep{dbr15}.
This interpretation is compatible with recent shock-structured and relativistic-flow models,
in which the emission region is extended and inhomogeneous, with multiple particle-acceleration
and cooling zones contributing at different orbital phases \citep{hkr21}.

The leptonic model predicts 3--70~keV synchrotron fluxes from
the relativistic-Maxwellian (Syn1) and broken-power-law (Syn2)
electron populations of $1.72\times10^{-11}$ (SUPC) and
$4.26\times10^{-12}$~erg~cm$^{-2}$~s$^{-1}$ (INFC), compared with
\textit{NuSTAR} values of $(1.65\pm0.03)\times10^{-11}$ and
$(3.19\pm0.04)\times10^{-11}$~erg~cm$^{-2}$~s$^{-1}$
\citep{yke+21}. SUPC is consistent, whereas INFC is low by
$\sim7.5$.

Although the purely leptonic calculation reproduces the main
GeV--TeV spectral structure, it underpredicts the $30$--$500$~GeV
emission, particularly during INFC, and the HAWC flux above
$\sim100$~TeV. These residuals motivate testing an additional
high-energy component in the following subsection.

\subsection{Lepto-hadronic Fit}

We extend the purely leptonic calculation by introducing a BPL proton population.
Its pion-decay emission is calculated using standard parameterizations of high-energy
$pp$ interactions \citep{rtk+03,kab06}, as described in Section~\ref{sec:rapa}.
The best-fit lepto-hadronic models are shown in Figure~\ref{fig:sedfit} (c) and (d),
and the corresponding parameters are listed in Table~\ref{tab:result}. In both phase
intervals, the emission below $\sim10$~GeV remains dominated by the relativistic-Maxwellian
and magnetospheric components, while shock-accelerated electrons produce the higher-energy
inverse-Compton emission extending through the TeV band. The additional
proton population contributes primarily in the multi-TeV to UHE regime.
In SUPC, $\alpha_{p,1}=2.08$ and $\log(E_{p,\rm break}/{\rm TeV})=2.43$, and the BIC decreases from
93.42 to 82.62 ($\Delta{\rm BIC}=10.80$). In INFC,
$\alpha_{p,1}=1.33$ and
$\log(E_{p,\rm break}/{\rm TeV})=1.07$, with the BIC decreasing from
140.54 to 136.27 ($\Delta{\rm BIC}=4.27$). These values indicate
stronger relative support for the lepto-hadronic description in SUPC,
but the reduced $\chi^2$ remains well above unity in both intervals
and the proton distributions remain weakly constrained. The BIC
differences therefore represent relative model preferences rather
than evidence for a uniquely adequate physical model.

For the modeled 10~TeV--0.346~PeV proton population,
$W_p=2.49\times10^{40}$~erg (SUPC) and
$9.06\times10^{40}$~erg (INFC). For
$t_{\rm res}=R_{\rm p}/c$, the residence times are 12.8 and 15.5~s
and $\langle f_{pp}\rangle=4.17\times10^{-6}$ and
$3.31\times10^{-6}$, consistent with rapid proton escape
\citep{bs13}. The required powers are
$2.08\times10^{39}$ and $5.99\times10^{39}$~erg~s$^{-1}$,
or $1.04\times10^3$ and $3.00\times10^3L_{\rm sd}$.
Thus the fast-escape local IBS-apex interpretation is energetically
inconsistent with pulsar spin-down alone. As an independent X-ray consistency check,
the lepto-hadronic fits predict 3--70~keV synchrotron fluxes of
$1.65\times10^{-11}$ (SUPC) and
$1.53\times10^{-12}$~erg~cm$^{-2}$~s$^{-1}$ (INFC).
The SUPC value agrees with \textit{NuSTAR}, whereas INFC is low by
$\sim21$ \citep{yke+21}. Thus, adding the proton component improves
the $\gamma$-ray description but does not resolve the INFC X-ray deficit.

The relative BIC support for the lepto-hadronic description is
stronger in SUPC, but the uncertainties do not establish an orbital
change in the proton spectral shape. The observed multi-TeV
modulation nevertheless indicates that at least part of the VHE
emission originates within or near the binary, where anisotropic IC
and $\gamma\gamma$ absorption remain important \citep{bk25}.
Future phase-resolved observations with VLAST \citep{fcg+22}, HAWC,
LHAASO, and CTAO will be required to constrain the origin of the
highest-energy component.

\subsection{Physical Implications for the UHE Emission}

The inferred phase dependence is broadly compatible with
spatially resolved transport calculations of LS~5039
\citep{hkr21,khg26}, but the present one-zone fits cannot distinguish
among variations in acceleration, escape, target density, and shock
geometry. Hadronic or hybrid interpretations have been considered in
pulsar-powered systems \citep{rtk+03,kab06,lcz10}, and strong
large-scale stellar-wind magnetic fields may in principle support
multi-PeV proton acceleration \citep{pb26}.

The electron-energy constraint applies specifically to the local
IBS-apex model: a $\sim100$~TeV IC photon requires an electron energy
of order $120$~TeV \citep{bk25}, well above the fitted
$28$--$41$~TeV limits. Relativistic outflows or larger-scale
termination shocks could relax the cooling and confinement constraints
\citep{bk25,mb26}, while magnetar-like activity could provide
additional nonthermal power \citep{bb25}. However, a steady
large-scale component alone would not explain the observed
$2$--$118$~TeV orbital modulation \citep{aaa+25}. Thus, the present
results constrain the local one-zone leptonic model but do not exclude
more complex leptonic origins of the UHE emission.
UHE emission from LS~I~+61$^\circ$~303 and the PWN powered by
PSR~J1849--0001 shows that $\gamma$-ray binaries and pulsar-powered
systems can accelerate particles to $>100$~TeV or PeV energies
\citep{cab+26,LHHAASO26}. Although these systems differ from LS~5039,
they demonstrate that such extreme particle energies are plausible in
compact pulsar-powered environments.

The statistical improvement obtained by including the proton
component favors a mixed phenomenological description of the broadband
SED. However, it does not uniquely identify the physical origin of the
UHE emission because the proton distribution remains weakly constrained
and extreme leptonic scenarios cannot be excluded. Additional
phase-resolved UHE and neutrino observations will be required to
distinguish between these possibilities.

\section{Summary}
\label{sec:sum}

We performed a $\gamma$-ray spectral analysis of LS~5039 using
17.3~yr of \textit{Fermi}-LAT data and identified an apparent
spectral excess at $\sim30$--$500$~GeV, with a local significance of $\sim7.90\sigma$.
We then constructed phase-resolved broadband SEDs by combining the
\textit{Fermi}-LAT results with published H.E.S.S.\ and HAWC
measurements and integrating the model over the full SUPC and INFC
intervals. The spectra show phase-dependent structure that motivates
a multicomponent interpretation, with the INFC SED exhibiting greater
spectral complexity than the SUPC SED.

Within the adopted framework, the GeV--TeV continuum is described
predominantly by leptonic components: the relativistic-Maxwellian and
magnetospheric components dominate below $\sim10$~GeV, while
shock-accelerated electrons produce IC emission above $\sim10$~GeV
and connect the GeV and TeV bands. However, the purely leptonic model
leaves residual structure at $30$--$500$~GeV and underpredicts the
HAWC flux above $\sim100$~TeV. At the highest energies, synchrotron
cooling limits the electron energies and IC scattering occurs in the
deep Klein--Nishina regime, making the UHE emission difficult to
reproduce within the local one-zone IBS-apex model, although more
complex leptonic acceleration sites or outflow geometries remain
possible.

Adding a proton component improves the broadband spectral description
in both phase intervals, reducing the BIC from 93.42 to 82.62 in SUPC
and from 140.54 to 136.27 in INFC. Thus, the relative support for the
lepto-hadronic model is substantially stronger in SUPC, although the
reduced $\chi^2$ remains well above unity in both intervals. These
results support an additional proton-induced contribution to the
high-energy emission whose relative strength or spectral properties
may vary over the orbit, although the fitted proton spectra remain
weakly constrained, particularly during INFC. The local IBS-apex
hadronic interpretation also remains energetically challenging under
the adopted fast-escape conditions.

Overall, our results are consistent with a multicomponent,
phase-dependent radiation scenario for LS~5039. Within the adopted
framework, an additional component improves the description of the
apparent $30$--$500$~GeV excess and the highest-energy emission.
However, its hadronic origin is not uniquely established because the
proton spectrum remains weakly constrained, the energy budget is
challenging, and more complex leptonic scenarios remain possible.
Future phase-resolved observations from $\sim10$~TeV to beyond
$100$~TeV with HAWC, LHAASO, and CTAO, complemented by
GeV-to-multi-TeV measurements with VLAST \citep{fcg+22} and searches for
phase-dependent neutrino emission, will help constrain its orbital
behavior and physical origin.

\begin{acknowledgments}
We would like to thank the anonymous referee for helpful suggestions.
This work was supported in part by the National Natural Science Foundation of China under Grant
No.~12233006. P.Z. acknowledges support from the Xingdian Talent Support Plan--Youth Project,
and by the Scientific Research and Innovation Project of Postgraduate Students in the Academic Degree of Yunnan University Nos.KC-262614856 and KC-262615807.
\end{acknowledgments}

\bibliographystyle{aasjournalv7}
%\bibliography{ref}
\bibliography{aas}

\end{document}